\documentclass[a4paper,11pt]{article}
\pdfoutput=1 % if your are submitting a pdflatex (i.e. if you have
\usepackage{jheppub} % for details on the use of the package, please
\usepackage[T1]{fontenc} % if needed

\usepackage[svgnames]{xcolor}
\usepackage{parskip}

\usepackage{breqn}
\allowdisplaybreaks

\usepackage{slashed}
\usepackage{bm}

\newcommand{\be}{\begin{equation}\begin{aligned}}
\newcommand{\ee}{\end{aligned}\end{equation}}
\newcommand{\nn}{\nonumber}
\newcommand{\0}{(0)}
\newcommand{\del}{\partial}
\renewcommand{\t}{\tilde}
\newcommand{\B}{\mathbf{B}}
\newcommand{\Del}{\nabla}

\title{\boldmath Dimensional Reduction and K\"ahler Metric for Metric Moduli in Imaginary Self-Dual Flux}

\author{Andrew R. Frey}
\author{and Ratul Mahanta}

\affiliation{\small Department of Physics and Winnipeg Institute for Theoretical Physics,
University of Winnipeg, 515 Portage Avenue, Winnipeg, Manitoba R3B 2E9, Canada}

\emailAdd{a.frey@uwinnipeg.ca}
\emailAdd{r.mahanta@uwinnipeg.ca}

\abstract{Understanding which effective field theories are consistent with an ultraviolet completion in quantum gravity is an important theoretical question. Therefore, it is important to know the structure of the 4D effective theory associated with a given compactification of string theory. 
We present a first-principles derivation of the low-energy 4D effective theory of geometric moduli in a warped Calabi-Yau compactification of type IIB string theory with imaginary self-dual 3-form flux. This completes the derivation of the metric on K\"ahler moduli space from the 10D equations of motion. We also give the first derivation of an effective action for flat directions in the complex structure moduli space of the Calabi-Yau (which generically mix with the axiodilaton) and work out explicit examples of complex structure flat directions in toroidal compactifications.
Finally, we outline applications to a variety of settings, including precision string phenomenology and the tadpole conjecture.}

\begin{document} 
\maketitle
\flushbottom

\section{Introduction}
\label{sec:intro}

Type IIB supergravity (SUGRA), in the presence of D-branes and O-planes, admits the Giddings-Kachru-Polchinski (GKP) backgrounds \cite{Giddings:2001yu}. These solutions feature a 10-dimensio\-nal warped product spacetime, where the 4D component is a warped Minkowski spacetime, and the six-dimensional internal manifold is a conformally Calabi-Yau (CY) orientifold. The warping is sourced by the 3-form flux $G_3$ which features components only in the compact directions. Moreover, $G_3$ is imaginary self-dual (ISD) with respect to the CY orientifold, i.e., $\tilde{\star}G_3=iG_3$, where $\tilde{\star}$ is the Hodge star operator associated with the CY orientifold. 
(The supersymmetric versions of GKP compactifications and similar compactifications of M-theory were discussed earlier in \cite{Becker:1996gj,Dasgupta:1999ss,Greene:2000gh}, and \cite{Becker:2001pm} described the M theory compactifications without reference to supersymmetry.)
The CY manifold has two types of deformations that preserve its Ricci flatness, which we will call geometric moduli: K\"ahler moduli (which are characterized by harmonic $(1,1)$ forms on the CY) and complex structure moduli (which are characterized by harmonic $(2,1)$ and $(1,2)$ forms on the CY). 
The K\"ahler deformations are moduli of the full 10D GKP background.\footnote{for a general CY compactification; the situation is slightly different when the internal manifold contains a torus factor.} 
Many of the complex structure deformations violate the ISD condition on the $G_3$ flux and are therefore stabilized, but some can remain moduli (possibly paired with a deformation of the axiodilaton).

Because GKP compactifications contain a classical mechanism for stabilizing some of the geometric moduli, they are an important testing ground for ideas in string phenomenology. Notably, with string ($\alpha'$) and nonperturbative contributions included, they have been proposed as a basis for de Sitter backgrounds in string theory \cite{Kachru:2003aw,vonGersdorff:2005bf,Berg:2005yu,Berg:2005ja,Balasubramanian:2005zx} (though not without controversy; see for example \cite{Sethi:2017phn,Kachru:2018aqn,Danielsson:2018ztv,Obied:2018sgi}) and further explorations in string cosmology. See \cite{Grana:2005jc,McAllister:2023vgy,Cicoli:2023opf} as reviews of many aspects of physics based on or inspired by GKP compactifications and their generalizations. 
Since some GKP compactifications are supersymmetric (and therefore expected to survive $\alpha'$ and quantum corrections), they are also key test cases for the swampland program, which asks about general properties that lower-dimensional effective field theories must possess in order to have an ultraviolet completion in a theory of quantum gravity such as string theory. For reviews of the swampland program, see \cite{Agmon:2022thq,VanRiet:2023pnx}.

In both contexts of string phenomenology/cosmology and the swampland program, the effective action of classically unstabilized moduli in the 4D effective theory, specifically the metric on moduli space, 
is of key importance. In the large volume limit, the warp factor becomes negligible and the flux dilute, so the effective action of a product compactification of Minkowski spacetime with a CY manifold (subject to an orientifold projection) is a good approximation \cite{Grimm:2004uq}; corrections to the moduli space metric due to the warp factor and flux are $\alpha'$ suppressed. 
However, strongly warped regions can appear even in regimes where curvatures remain small, so the full structure of the GKP compactification including warp factor and flux plays a role in determining the effective action. 

There have so far been two approaches used to derive the moduli space metric (or equivalently the K\"ahler potential of the 4D SUGRA). 
One, followed by \cite{Martucci:2009sf,Martucci:2014ska,Martucci:2016pzt}, uses combined considerations of the expected $N=1$ supersymmetry (SUSY) properties of the 4D effective theory and the structure of the 10D vacua, to determine the holomorphic coordinates and K\"ahler potential of the 4D SUGRA. The K\"ahler potential for the K\"ahler moduli (geometric moduli and axionic partners), 2-form moduli, and D3-brane moduli, appeared in \cite{Martucci:2016pzt}. 
The other approach is a direct dimensional reduction by finding massless solutions to the 10D equations of motion. 
As pointed out initially by \cite{Frey:2002hf,Frey:2003tf,Giddings:2005ff}, the equations of motion include nontrivial constraints due to gauge invariance, which can be solved by introducing additional nonvanishing components of the 10D fields known as compensators; \cite{Douglas:2008jx,Shiu:2008ry,Underwood:2010pm} demonstrated the use of different gauge choices in the dimensional reduction. 
This first-principles approach has been applied to the universal volume (K\"ahler) modulus \cite{Frey:2008xw}, axions descending from $C_4$, $C_2$, and $B_2$ form fields \cite{Frey:2013bha}, and $D3$-brane position moduli \cite{Cownden:2016hpf}. The $C_4$ axions are supersymmetric partners of the geometric K\"ahler moduli; these moduli together share a single moduli space metric and are often together known as K\"ahler moduli of the 4D SUGRA.\footnote{An apparently related approach to dimensional reduction for heterotic string compactifications appears in \cite{Candelas:2016usb,McOrist:2016cfl,McOrist:2025sdy}.}

In this article, we address the case of unstabilized geometric moduli from first principles. 
To ensure that the deformed 10D system remains in the moduli space of the GKP compactification, we restrict $G_3$ (in the compact directions) to remain ISD with respect to the deformed CY metric. Following \cite{Martucci:2016pzt}, we will see that this is possible for all geometric K\"ahler moduli, although generically $G_3$ must change in response to the change of metric. 
While generic geometric complex structure moduli cannot satisfy the ISD condition, some do, possibly in tandem with a deformation of the axiodilaton. We will refer to these unstabilized moduli as complex structure flat directions whether they involve the axiodilaton or not.
The field space metric for all these moduli is phenomenologically important.
Since they are classically massless, K\"ahler moduli have long been of interest in cosmology, for example, in models of inflation (with one or several inflaton fields). 
Complex structure flat directions are of more recent interest. The tadpole conjecture suggests that many of the complex structure moduli are flat \cite{Bena:2020xrh, Bena:2021wyr, Grana:2022dfw, Lust:2022mhk, Lust:2021xds, Plauschinn:2021hkp}. Such flat directions, when stabilized with non-perturbative complex structure instanton corrections to the prepotential, have been used to obtain an exponentially low stabilized value of the superpotential \cite{Demirtas:2019sip,Broeckel:2021uty}. For specific cases like toriodal orientifold or specific CY orientifolds, explicit flux vacua with complex structure flat directions are systematically classified in \cite{Cicoli:2022vny} generalizing ideas first appearing in \cite{Hebecker:2017lxm}.

For the K\"ahler moduli, this work bridges a gap in the literature by deriving the 4D kinetic action through dimensional reduction, completing the derivation of the moduli space metric for all the K\"ahler moduli.
We also provide the kinetic action and moduli space metric of unstabilized complex structure moduli for the first time in the literature.
Because our methodology requires only the equations of motion, they apply whether the GKP background breaks SUSY or not, though we will point out how SUSY breaking affects the interpretation of our results.
Similarly, since the flux and warp factor appear at lower order in $\alpha'$ than string corrections to the SUGRA (as we describe in section \ref{sec:gkpbkgd} following \cite{Becker:2001pm}), we work with the classical 10D SUGRA and are not concerned with whether the GKP background survives corrections to the SUGRA.

The article is organized as follows. Section \ref{sec:Review} provides a review of the necessary details of the GKP background, outlines the general framework for dimensional reduction to derive the 4D effective action from the 10D action, and reviews the behavior of products of harmonic forms. 
Section \ref{sec:GeoModISD} explores the geometric moduli space (K\"ahler and flat complex structure) of GKP solutions in keeping with the ISD 3-form flux $G_3$ (we showcase explicit examples of complex structure flat directions on toroidal orientifolds in appendix \ref{app:CompStrFlat}). 
In section \ref{s:linearization}, we present our ansatz for linear fluctuations of the metric and the form fields and show how to solve the linearized 10D SUGRA equations of motion and Bianchi identities. Specifically, section \ref{sec:Constr.dynEOM} derives the constraints on the linear fluctuations as well as the dynamical equations of motion, while section \ref{sec:SolveConstr} focuses on solving these constraints. In section \ref{sec:ModuliActionMetric}, we derive the kinetic action for the Kähler and flat complex structure moduli, ensuring the constraints are satisfied off-shell. Finally, section \ref{sec:Discuss} concludes the paper and discusses potential applications of our results.

Although appendix \ref{app:conventions} compiles all the conventions used throughout this paper, we highlight the following to help the reader. A tilde on upper indices denotes those raised by the internal CY metric of the GKP background, whereas a hat on upper indices indicates those raised by the 4D Minkowski metric. $\hat d$ is the external derivative with respect to noncompact spacetime coordinates, $\tilde{d}$ denotes a differential with respect to  internal coordinates, and covariant derivatives with tilde are defined using the Christoffel connection for the background CY. An $[i,j]$-form refers to a differential form on 10D spacetime with $i \leq 4$ non-compact and $j \leq 6$ compact indices. Meanwhile, an $(i,j)$-form on the internal complex manifold specifies a differential form with $i\leq3$ holomorphic and $j\leq3$ antiholomorphic indices. 

%\rma{A possible list of papers for citations: \cite{Frey:2013bha, Cownden:2016hpf, Giddings:2001yu, Martucci:2016pzt, Polchinski:1998rr, Johnson:2000ch, Kachru:2003aw, Grimm:2004uq, Grana:2005jc, Lust:2022xoq, Cicoli:2022vny, Underwood:2010pm, Frey:2008xw}.}

%\rma{New references: \cite{Giddings:2005ff, Frey:2002hf, Koerber:2007xk, Martucci:2009sf, Martucci:2014ska}.}

\section{Review}
\label{sec:Review}

In this section, we review the Giddings-Kachru-Polchinski (GKP) background in type IIB supergravity, around which linear fluctuations are considered in this article. Subsequently, we briefly outline the dimensional reduction methodology in deriving moduli action for later use
and some results on the products of harmonic forms.

\subsection{Giddings-Kachru-Polchinski background}
\label{sec:gkpbkgd}

The GKP background is a solution to the type IIB SUGRA equations of motion and Bianchi identities given in \eqref{eq:eomforms2}-\eqref{eq:eomeinstein}, where the fields take the following form:
\begin{align}
  &ds^2=e^{2\Omega} e^{2A(y)}\hat{\eta}_{\mu\nu}dx^{\mu}dx^{\nu}+e^{-2A(y)}\tilde{g}_{mn}(y)dy^mdy^n\,,\nn\\
  &\tilde{F}_5=e^{4\Omega}\hat{\epsilon}\wedge\tilde{d}e^{4A(y)}+\tilde{\star}\tilde{d}e^{-4A(y)}\,,\quad \tilde{\star}G_3(y)=iG_3(y)\,, \label{eq:gkpbg}
\end{align}
where $x^\mu,y^m$ are respectively the coordinates along the 4D spacetime and the 6D internal manifold. $\hat{\eta}_{\mu\nu}$ is the Minkowski metric, and $\tilde{g}_{mn}$ a Calabi-Yau (CY) metric (therefore Ricci flat). 
These are orientifold compactifications with either O3- or O7-planes as well as corresponding D-branes.
For simplicity, we assume that any D7-brane charge cancels locally (i.e., we work in the orientifold limit of F theory), so the axiodilaton $\tau \equiv C_0+ie^{-\Phi}$ takes a constant background value. 
The complex 3-form is defined $G_3\equiv F_3 -\tau H_3$; the real 3-forms are locally defined $F_3=dC_2$ and $H_3=dB_2$.
The metric, $\tau$, and $C_4$ have positive parity under the orientifold involution, while the 2-form potentials (and $G_3$) have negative involution parity.
In addition to the ISD condition on $G_3$, the $(0,3)$ component of $G_3$ vanish in terms of the complex coordinates of the CY if the background preserves SUSY.

Away from local sources, the warp factor $A$ satisfies a Poisson equation given by
\begin{align}
   \tilde{d}\tilde{\star}\tilde{d}e^{-4A(y)}=\frac{i}{2}\frac{G_3\wedge\bar{G}_3}{\textrm{Im}\tau}\,. \label{eq:gkpbgwarpeqn}
\end{align}
This equation has three important consequences. First, it is still satisfied if we add a constant to $e^{-4A}$; such a shift is an overall volume modulus for the compactification.
We define the Weyl factor as 
\be\label{eq:defOmega} e^{-2\Omega}\equiv 
\frac{1}{\tilde{V}} \int d^6y \sqrt{\tilde g} e^{-4A}\,,\quad \tilde{V}\equiv \int d^6y\sqrt{\tilde g}\,,\ee
so the 4D Planck scale is constant over this moduli space ($\hat\eta_{\mu\nu}$ is the Einstein frame metric). At any given point in moduli space, $\Omega$ is simply a constant. Second, scaling $\tilde{g}_{mn}$ by a constant also scales the warp factor such that the only change to the 10D metric is an overall scaling of the noncompact metric. This can be removed by rescaling the $x^\mu$ coordinates, so it is pure gauge. 
Finally, the integral of \eqref{eq:gkpbgwarpeqn} vanishes, so the total integral of $G_3\wedge \bar{G}_3$ matches the total of the local sources; this equation is the tadpole constraint.

Given the background fields \eqref{eq:gkpbg}, the equations of motion mean that $G_3$ is harmonic. 
As a result, the real 3-forms $F_3,H_3$ cannot be written globally in terms of 2-form potentials; furthermore, the fact that fundamental and D-strings are charged under these fields implies that the integrals of $F_3,H_3$ on any 3-cycle are quantized in units of $4\pi^2\alpha'$, where $1/2\pi\alpha'$ is the fundamental string tension.
The condition that the flux is ISD restricts the moduli of the metric since any deformation of the metric must maintain the ISD condition as well as keeping the metric Calabi-Yau (and flux quantization prevents the ISD condition from becoming trivial).
As pointed out by GKP and we will see in detail below, the K\"ahler moduli maintain the ISD condition, while the complex structure moduli and the axiodilaton generically do not.\footnote{The discussion is somewhat modified if the compactification contains a torus factor, to be discussed below.} 
However, the number of flux quanta of $G_3$ (including independent components) is limited by the tadpole constraint. The tadpole conjecture \cite{Bena:2020xrh, Bena:2021wyr, Grana:2022dfw, Lust:2022mhk, Lust:2021xds, Plauschinn:2021hkp} suggests that the number of flux quanta needed to stabilize all the complex structure moduli grows faster with the number of moduli than the tadpole constraint allows, so we expect that there are nonetheless flat directions in the complex structure moduli space as well in many GKP compactifications.

There are additional moduli of these backgrounds. The form potentials $C_4,C_2,B_2$ can take values proportional to harmonic forms on the CY (of the appropriate involution parity) without contributing to the field strengths, and the proportionality constants are moduli (commonly referred to as axions, though not all have shift symmetries). There are also moduli associated with the degrees of freedom of any D-branes in the background.

Based on the structure of the 10D background, there are a few points we can make immediately about the 4D effective field theory. 
If the $(0,3)$ component of $G_3$ vanishes, the background is supersymmetric, so the 4D theory must be described by SUGRA. 
Even in the case that SUSY is broken, the supersymmetry breaking scale is hierarchically smaller than the Kaluza-Klein scale for large compactification volume, so the 4D theory is still a SUGRA but with spontaneous SUSY breaking. 
Away from large volume, including compactifications with strong warping, the scales are not strongly separated, but the effective theory is still assumed to be 4D SUGRA.

The reader may have noticed that the 3-form quantization means that any derivative of the warp factor is formally of order $(\alpha')^2$. 
Before proceeding, we make two comments. First, it is reasonable to expect --- and evidence strongly suggests \cite{McAllister:2024lnt} --- that there exist compactifications with strongly warped regions (near conifold points, for example) that are nonetheless weakly curved everywhere.
In that case, the 10D geometry is not a perturbation around a product compactification with a CY manifold.
Second, in the large volume limit, the warp factor is approximately constant with $\t\nabla^{\t 2} A\sim |G_3|^{\t 2}\sim (\alpha')^2$.\footnote{This is also the same order as the contribution of D3- and D7-branes and O3- and O7-planes to the Poisson equation \eqref{eq:gkpbgwarpeqn} for the warp factor.} 
While the $(\del A)^{\t 2}$ terms in $\t\nabla^{\t 2} e^{-4A}$ are of the same order as $(\alpha')^4$ corrections to the supergravity action, the warp factor and flux nonetheless appear in the classical 10D SUGRA, i.e., consistently at a lower order in $\alpha'$.
Therefore, the GKP background is a consistent solution to the classical SUGRA, regardless of SUSY, which depends on the decomposition of $G_3$ into components in terms of complex coordinates. This is the same as the argument given by \cite{Becker:2001pm} for M theory compactifications.
Because our methodology does not depend on the components of $G_3$, we will work at leading order and do not address the question of whether higher-order corrections destabilize the compactification in the non-supersymmetric case \cite{Sethi:2017phn} or whether nonperturbative corrections can stabilize it again \cite{Kachru:2018aqn}.

Henceforth, we will explore fluctuations around a single fixed background of this type, moving to a second GKP background with the same topological quantities (ie, the same CY manifold and the same integrals of the background $G_3$).
We denote the field values of the initial fixed background with a subscript or superscript $\0$.

\subsection{Dimensional reduction scheme}
\label{sec:review.dim.reduce}

Now we outline the general scheme of dimensional reduction for a first-principles derivation of the 4D effective theory from the 10D supergravity, for later use in the case of a geometric modulus.

Schematically, consider an action functional in 10D that depends on the fields $\psi_a$ and their derivatives up to second order. The action and the corresponding Euler-Lagrange equations of motion are given by
\begin{align}
  S=\int d^4xd^6y\mathcal{L}(\psi_a,\partial_M\psi_a,\partial_M\partial_N\psi_a)\,, \quad E^a\equiv\partial_M\partial_N\frac{\delta\mathcal{L}}{\delta\partial_M\partial_N\psi_a}-\partial_M\frac{\delta\mathcal{L}}{\delta\partial_M\psi_a}+\frac{\delta\mathcal{L}}{\delta\psi_a}=0\,,
\end{align}
where $\mathcal{L}$ includes $\sqrt{-g}$. We refer to $E^a$ as the equation of motion (EOM) operators. Suppose $\psi_a^{(0)}$ are background fields that satisfy above equations of motion, i.e. $E^a_{(0)}=0$. Linear fluctuations around $\psi_a^{(0)}$ result in linear fluctuations in $E^a$, as follows:
\begin{equation}
  \psi_a=\psi_a^{(0)}+\delta\psi_a\,,\quad E^a=E^a_{(0)}+\delta E^a\,,
\end{equation}
where $\delta E^a$ involve the fluctuations $\delta\psi_a$ and their derivatives $\partial_M\delta \psi_a,\partial_M\partial_N\delta\psi_a$. At second order, the action can be organized as
\begin{align}
  S=\frac{1}{2}\int d^4xd^6y\, \delta\psi_a\delta E^a\,.
\end{align}
At zeroth order, we get a constant $S_{(0)}$ obtained by evaluating the action at the background fields which we ignore. The linear order terms cancel, due to the background equations of motion. One can also keep second order variations around the background fields, and prove that those do not contribute to the action at the second order. For further details, see \cite{Shiu:2008ry,Frey:2013bha}.

For the case of type IIB supergravity the EOM operators can be taken as
\begin{align}\label{eq:EOMoperators}
  &E_{MN}\equiv G_{MN}-T_{MN}\,, \nn\\
  &E_{\tau}\equiv d\star d\tau+\frac{i d\tau\wedge\star d\tau}{\textrm{Im} \tau}+\frac{i}{2}G_3\wedge\star G_3\,, \nn\\
  &E_6\equiv d\star\tilde{F}_5-\frac{i G_3\wedge\bar{G}_3}{2\textrm{Im}\tau}\,, \nn\\
  &E_8\equiv d\star G_3-\frac{i}{2}(\tilde{F}_5-\star\tilde{F}_5)\wedge G_3+\frac{i d\tau\wedge\star\textrm{Re}G_3}{\textrm{Im}\tau}\,,
\end{align}
and we define $\bar{E}_{\bar{\tau}}$ and $\bar{E}_8$ as complex conjugates to $E_\tau$ and $E_8$, respectively. In terms of the linear variations of these EOM operators the quadratic action can be written as
\begin{align}\label{eq:EFTaction}
  S=&\frac{1}{4\kappa^2}\bigg[\int d^{10}X\sqrt{-g}\delta g^{MN}\delta E_{MN}-\frac{1}{2\left(\textrm{Im}\tau^{(0)}\right)^2}\int\left(\delta\tau\delta\bar{E}_{\bar{\tau}}+\delta\bar{\tau}\delta E_\tau\right) \nn\\
  &\qquad\qquad -\int \delta C_4^\prime\wedge\delta E_6-\frac{1}{2\textrm{Im}\tau^{(0)}}\int \left(\delta A_2\wedge\delta\bar{E}_8+\delta\bar{A}_2\wedge\delta E_8\right)\bigg]\,,
\end{align}
where $\delta A_2\equiv \delta C_2-\tau^{(0)}\delta B_2$ and $\delta C_4^\prime\equiv \delta C_4-C_2^{(0)}\wedge\delta B_2/2+B_2^{(0)}\wedge\delta C_2/2$. 
As discussed in \cite{Frey:2002hf,Frey:2013bha,Cownden:2016hpf}, the fluctuation $\delta C_4$ is not globally defined because the background $C_2^\0$ and $B_2^\0$ are also not globally defined --- if they were, $G_3^\0$ would not be harmonic. Instead, the redefined $\delta C'_4$ is globally defined (it does not transform under the gauge transformations of $C_2^\0$ and $B_2^\0$) and is suitable to describe a 4D degree of freedom. This field redefinition also removes an additional term (which we have already omitted from \eqref{eq:EOMoperators}) from the EOM operator $E_8$. 
There is another subtlety related to the self-duality of $\t F_5$; namely, the components of $\delta C'_4$ (or $E_6$) are not independent. 
In fact, if we included all the components of $\delta C'_4$ (and $\delta E_6$) in \eqref{eq:EFTaction}, the corresponding term would vanish. 
One approach, as stated in \cite{Giddings:2001yu}, is to keep only a set of independent components of $\delta C'_4$. Following \cite{Cownden:2016hpf}, we keep only the ``magnetic'' $[0,4]$ and $[1,3]$ components ($[2,2]$ components that are $\hat d$ exact are also magnetic but do not appear in this work). Restricting to only the magnetic components also requires us to double the action in the 5-form sector, which we have already done in \eqref{eq:EFTaction}. For more rigorous approaches to the IIB SUGRA action, see \cite{arXiv:hep-th/9806140,Frey:2019,Evnin:2022kqn}.

Because the action is an off-shell quantity, we should not set the EOM operators $\delta E^a$ to zero, which amounts to satisfying the 10D supergravity equations of motion. 
However, because the compactification breaks part of the gauge and diffeomorphism invariance, some components of $\delta E^a$ are constraints that must vanish. The action \eqref{eq:EFTaction} describes four-dimensional degrees of freedom after integration over the compact dimensions, but it is only invariant under the full gauge and diffeomorphism transformations of the 10D spacetime if the constraints are satisfied. (See \cite{Frey:2013bha} for a further discussion and example.)
These constraints have overall multiplicative factors $\partial_\mu u$ and $\partial_{\mu}\partial_{\nu}u$ (with $\mu\neq\nu$) and determine the $y$ dependence of the fluctuations, and we impose the constraints off-shell to simplify and perform the $y$-integrations to evaluate the 4D kinetic action for the modulus $u(x)$.
The other equations of motion are dynamical equations of motion in the form of a Klein-Gordon equation with terms proportional to $u$ and its d'Alembertian. In this paper, we also restrict to the moduli space of the compactification, which is another constraint setting the terms proportional to $u$ equal to zero. We also impose this constraint off-shell, ensuring that we are working with massless modes. 
After imposing this constraint, the dynamical equation of motion is a massless Klein-Gordon equation, which \emph{we do not satisfy off-shell.}

%This yields four types of equations involving the modulus $u(x)$ and its derivatives. The first and second types respectively have overall multiplication factors $\partial_\mu u$ and $\partial_{\mu}\partial_{\nu}u$ (with $\mu\neq\nu$), while the third and fourth types respectively involve $u$ and $\partial^{\hat{2}}u$ or only $\partial^{\hat{2}}u$. For the third type of equation, we set the $u$ terms to zero to ensure we are working with massless modes, i.e., staying on the moduli space of the 10D equations of motion (which can be obtained when the $x$-dependency of $u$ is dropped). This requirement and the first two types of equations result in {\it constraints} on the $y$-dependencies of the fluctuations. With this, the third and fourth types equations lead to the Klein-Gordon equation $\partial^{\hat{2}}u=0$, which we refer as the {\it dynamical equation and do not satisfy off-shell}. After obtaining the quadratic action mentioned above, we impose the constraints off-shell to simplify and perform the $y$-integrations to evaluate the 4D kinetic action for the modulus $u(x)$.

In the following sections, we additionally use a subscript $u$ on the variations $\delta$ to emphasize the dependence on the infinitesimal modulus $u$. The variation $\delta$ without this subscript, which will appear more frequently, refers to the coefficients of the fluctuations when the dependencies on $u$ are factored out.

\subsection{Products of harmonic forms} \label{s:harmonicproducts}

The effective action \eqref{eq:EFTaction} will contain various products of harmonic forms on the CY (specifically forms that are harmonic with respect to the background metric $\t g_{mn}^\0$). These arise directly in the contributions from the form EOM operators and also in the contribution from the Einstein equation operator because the geometric moduli of the Ricci-flat metric can be written in terms of harmonic forms \cite{Candelas:1990pi}. Here we review some properties of products of harmonic forms for general metrics.

Consider first the pointwise inner product of two harmonic $p$-forms, $\alpha_p\cdot\beta_p$ as defined in appendix \ref{app:conventions}, which is generally a function on the CY manifold.\footnote{The $L^2$ inner product of the forms is the integral of the pointwise inner product.} 
An important question in determining the effective action of geometric moduli in warped compactifications is whether this pointwise inner product is constant (this question is also important for axionic moduli \cite{Frey:2013bha}). 
This question is also of interest in pure mathematics \cite{nagy2004,nagy2006,grosjean}, and the answer is that generally the pointwise inner product is \emph{not} constant for two harmonic forms, nor do the special properties of K\"ahler or CY metrics make it constant. It is in fact impossible for all pointwise inner products of two harmonic $p$-forms on an $n$-dimensional (Euclidean) manifold to be constant if the real Betti number $b_p$ is larger than that of the torus $T^n$ \cite{kotschick2000}.
It is already possible to see that pointwise inner products of harmonic 2-forms are nonconstant on K3 \cite{kotschick2000}.

Up to sign, this pointwise inner product is $\t\star(\alpha_p\wedge\t\star\beta_p)$, which is automatically constant whenever $\alpha_p\wedge\t\star\beta_p$ is harmonic. We can therefore also consider the more restrictive conditions such that all wedge products of harmonic forms are themselves harmonic, which occurs for \textit{formal} metrics. 
Based on our above discussion, we see that the Ricci-flat metric on a CY manifold is generally \emph{not} formal, so the wedge product of harmonic forms need not be harmonic; CY manifolds in general do not admit any formal metric.
It is true that all K\"ahler manifolds satisfy some necessary topological conditions to admit a formal metric \cite{deligne}; those conditions are necessary but insufficient. The only formal metric on a torus is the flat metric \cite{kotschick2000}.

There is one important special case to consider.
In terms of complex coordinates $z^i,\bar z^{\bar\imath}$ on the internal CY manifold, the metric is Hermitian with one holomorphic and one antiholomorphic index, given by $\tilde{g}_{i\bar{\jmath}}$, while $\tilde{g}_{ij} = \tilde{g}_{\bar{\imath}\bar{\jmath}}=0$. This metric is associated with the harmonic K\"ahler form $\tilde{J}_2$ with $\tilde{J}_{i\bar{\jmath}}=i\tilde{g}_{i\bar{\jmath}}$. 
On any K\"ahler manifold, the wedge product of $\t J_2$ on any form commutes with application of the Hodge-de Rham Laplacian $\triangle$ associated with the metric $\tilde{g}_{i\bar{\jmath}}$. As a result, the wedge product of $\t J_2$ with any harmonic form is itself harmonic; similarly, the contraction of $\t J_2$ with any harmonic form is a harmonic form (so the pointwise inner product of $\t J_2$ with any harmonic 2-form is constant).

\section{Geometric moduli in ISD flux}
\label{sec:GeoModISD}

We consider linear fluctuations $\delta_u\t g_{mn}\equiv u\delta\tilde{g}_{mn}(y)$ (where $u$ is infinitesimal) of the unwarped CY metric components $\tilde{g}^{(0)}_{mn}(y)$, ensuring that it remains Ricci flat and preserves the unwarped volume at linear order. Therefore we have\footnote{Tildes on covariant derivatives $\tilde{\nabla}_p$ and differentials $\tilde{d}$ indicate those w.r.t. $y$-coordinates (using the Christoffel connection for $\tilde{g}^{(0)}_{mn}$ in the case of $\tilde{\nabla}_p$). Tilde on upper indices indicate raises by the background inverse metric $\tilde{g}^{(0)}{}^{pl}$.}
\begin{align}
  &\tilde{\nabla}^{\tilde{l}}\tilde{\nabla}_{m}\delta\tilde{g}_{nl}+\tilde{\nabla}^{\tilde{l}}\tilde{\nabla}_{n}\delta\tilde{g}_{ml}-\tilde{\nabla}^{\tilde{l}}\tilde{\nabla}_{l}\delta\tilde{g}_{mn}\nn\\
  &\qquad=\tilde{\nabla}_{m}\tilde{\nabla}^{\tilde{l}}\delta\tilde{g}_{nl}+\tilde{\nabla}_{n}\tilde{\nabla}^{\tilde{l}}\delta\tilde{g}_{ml}-2R_{(0)}{}^{\tilde{l}}{}_m{}^{\tilde{p}}{}_n\delta \tilde{g}_{lp}-\tilde{\nabla}^{\tilde{l}}\tilde{\nabla}_{l}\delta\tilde{g}_{mn}=0\,, \nn\\
  &\tilde{g}^{(0)}{}^{mn}\delta\tilde{g}_{mn}=0\,, \label{eq:metricmodcons}
\end{align}
where $R_{(0)}{}_{lmpn}$ is the Riemann tensor w.r.t. the background $\tilde{g}^{(0)}_{mn}$, and we have used Ricci flatness of the background metric i.e. $R_{(0)}{}_{mn}=0$. 
It is also common to impose the condition $\tilde{\nabla}^{\tilde{n}}\delta\tilde{g}_{nm}=0$ (the metric deformation is covariantly transverse), which is a gauge choice for constant $u$. We will later see that this condition is required for varying $u(x)$.
Following \cite{Candelas:1990pi}, we will assume that the metric deformation is covariantly transverse in order to write $\delta\t g_{mn}$ in terms of harmonic forms as described in the following subsections.

We refer to $u$ as a \emph{geometric modulus}. However, as we described in section \ref{sec:gkpbkgd}, GKP backgrounds have additional requirements on the flux and warp factor beyond the CY condition on the metric. In this section, we find the conditions on linear fluctuations of the additional SUGRA fields that allow the geometric modulus $u$ to remain an (unstabilized) modulus of the full GKP compactification.
These are solutions to the first order equations of motion for static deformations, i.e., constant $u$.
For all fields, we will write $\delta_u\psi \equiv u\delta\psi(y)$. Since we eventually promote the modulus to a function $u(x)$ of the external spacetime, $\delta_u\psi$ is the full fluctuation in 10D, and $\delta\psi(y)$ has the $x$ dependence factored out.

\subsection{Static deformations}\label{s:staticmod}
One of the characteristic features of GKP compactifications is the ISD nature of the 3-form flux (as required for the background to solve the SUGRA EOM).
Therefore, for $u$ to be a modulus, the (compact direction components of) perturbed 3-form flux $G_3$ must satisfy the ISD condition with respect to the perturbed unwarped metric. That is, up to linear order we require\footnote{An $[i,j]$-form refers to a differential form on 10D spacetime with $i$ non-compact and $j$ compact indices.}
\begin{align}
  (\tilde{\star}^{(0)}+u\delta\tilde{\star})\left(G_3^\0+u\chi_3 \right)=i\left(G_3^\0+u\chi_3 \right)\, ,
\end{align}
where $\delta_u G_{[0,3]}\equiv u\chi_3(y)$ is the fluctuation in $G_3$. 
Therefore, the ISD condition links $\delta\tilde{g}$ to  $\chi_3$, through %\footnote{In obtaining the second equality, the ISD property of the background flux $G_3^{(0)}$ is utilized.}
$W_3=\chi_3+i\tilde{\star}^{(0)}\chi_3$,
where we define
\begin{align}
  \delta_u\tilde{\star}G_3^{(0)}\equiv -\tilde{\star}^{(0)}\delta_u W_3= -u\t\star^\0 W_3\,,\quad W_{mnp}\equiv 3\delta\tilde{g}_{l[m}G^{(0)}{}^{\tilde{l}}{}_{np]}\,. \label{eq:W3def}
\end{align}
From the definition of $W_3$ and the ISD property of the background 3-form $G_3^{(0)}$, it follows that $W_3$ is AISD
\begin{align}
  \tilde{\star}^{(0)}W_3=-iW_3\,, \label{eq:ISDW3}
\end{align}
which is consistent with its relationship to $\chi_3$. 

Since the 3-form flux $G_3$ is given by %\footnote{For further details see appendix \ref{app:IIBSUGRAeom}.}
$G_3=F_3-\tau H_3$, fluctuations $\delta_u C_2,\delta_u B_2,\delta_u\tau$ in the potentials induce a $[0,3]$ linear fluctuation %$\delta_u G_3$ around its harmonic background value $G_3^{(0)}$, given by
\begin{align}
  \delta_u G_3=d\delta_u A_2+\delta_u\tau\frac{G_3^{(0)}-\bar{G}_3^{(0)}}{2i\textrm{Im}\tau^{(0)}}\,,\quad \delta_u A_2\equiv\delta_u C_2-\tau^{(0)}\delta_u B_2\,. \label{eq:G3fluctuation}
\end{align}
Taking $\delta_uA_2=u\eta_2(y)$ and $\delta_u\tau=u\delta\tau$ with $\delta\tau$ assumed constant, the static deformation is %with constant $\delta\tau$. For the case of static deformations, where $u$ is treated as a constant parameter, we obtain
\begin{align}
  %\delta_u G_3=u\chi_3(y)\,,\quad 
  \chi_3(y)\equiv\tilde{d}\eta_2+\delta\tau\frac{G_3^{(0)}-\bar{G}_3^{(0)}}{2i\textrm{Im}\tau^{(0)}}\,. \label{eq:chi3def}
\end{align}
Clearly $\chi_3$ is closed as $G_3^{(0)}$ is harmonic.

The linearized ISD condition becomes
\be W_3 =\tilde{d}\eta_2+i\tilde{\star}^{(0)}\tilde{d}\eta_2+\frac{i\delta\tau}{\textrm{Im}\tau^{(0)}}\bar{G}_3^{(0)} \label{eq:ISDG3}
\ee
where the $G_3^\0$ term vanishes because the background flux is ISD.
Moreover, acting $\tilde{d}$ on \eqref{eq:ISDG3} results in
\begin{align}
  \tilde{d}\tilde{\star}^{(0)}\tilde{d}\eta_2=-i\tilde{d}W_3\,, \label{eq:Eta2EqnDirect}
\end{align}
due to the closedness of the background flux $G_3^{(0)}$. 
We impose $\t d\t\star^\0 \eta_2 =0$ (for constant $u$, this is a gauge choice, but we will find it is required for spacetime dependent fluctuations), so \eqref{eq:Eta2EqnDirect} becomes a Poisson equation.

Through \eqref{eq:gkpbgwarpeqn} and \eqref{eq:defOmega}, the variation of the metric introduces fluctuations $\delta_u A$ and $\delta_u\Omega$ in $A$ and $\Omega$. It is convenient to introduce the notation
\begin{align}
  \delta e^{-4A}\equiv -4e^{-4A^{(0)}}\delta A\,,\quad \delta e^{-2\Omega}\equiv -2e^{-2\Omega^{(0)}}\delta\Omega
\end{align}
for later use. Note that $\delta\Omega$ is constant in $y$.
The above fluctuations, in turn, generate a fluctuation in $\tilde{F}_5$ through the expansion of \eqref{eq:gkpbg}:
\begin{align}
  \delta_u\tilde{F}_{[0,5]}=-u\tilde{\star}^{(0)}V_1+u\tilde{\star}^{(0)}\tilde{d}\delta e^{-4A}\, .\label{eq:fluctF05}
\end{align}
We have defined
\begin{align}
  %\delta\tilde{\star}\tilde{d}e^{-4A^{(0)}(y)}\equiv -\tilde{\star}^{(0)}V_1(y)\,,\quad 
  V_{m}=\delta\tilde{g}_{mn}\partial^{\tilde{n}}e^{-4A^{(0)}}\,. \label{eq:V1def}
\end{align}
%
% We now discuss how linearized EOM and Biancchi identities are solved when $G_3$ flux remains ISD. We start with the $\tilde{F}_5$- Bianchi identity (which is same as $C_4$ equation of motion). Owing to above fluctuations this only acquires $[0,6]$ components proportional to $u$, at linear order. Equating them on both sides of the Bianchi identity, we get
% %
% \begin{align}
%   \tilde{d}\tilde{\star}^{(0)}\tilde{d}\delta e^{-4A}-\tilde{d}\tilde{\star}^{(0)}V_1=\frac{i}{2\textrm{Im}\tau^{(0)}}\left(G_3^{(0)}\wedge \bar{\chi}_3+\chi_3\wedge\bar{G}_3^{(0)}\right)-\frac{i\textrm{Im}\delta\tau}{2(\textrm{Im}\tau^{(0)})^2}G_3^{(0)}\wedge\bar{G}_3^{(0)}\,,
% \end{align}
% %
% Using the definition of $\chi_3$ in \eqref{eq:chi3def}, above constraint can be simplified further to
The linearized Bianchi identity becomes
\begin{align}
  \tilde{d}\tilde{\star}^{(0)}\tilde{d}\delta e^{-4A}-\tilde{d}\tilde{\star}^{(0)}V_1=\frac{i}{2\textrm{Im}\tau^{(0)}}\left(G_3^{(0)}\wedge \tilde{d}\bar{\eta}_2+\tilde{d}\eta_2\wedge\bar{G}_3^{(0)}\right)\,. \label{eq:F5Bianchi06}
\end{align}
Since $\delta A$ and $\delta\Omega$ appear in the metric, the $(\mu,\nu)$ component of the Einstein equation is similarly nontrivial:
\begin{align}
  &\frac{1}{2}e^{2\Omega^{(0)}+8A^{(0)}}\left(\tilde{\nabla}^{\tilde{2}}\delta e^{-4A}-\tilde{\nabla}^{\tilde{m}}V_{m}\right)\hat{\eta}_{\mu\nu} \nn\\
  &\quad=\frac{1}{4}\hat{\eta}_{\mu\nu}\bigg[\frac{e^{2\Omega^{(0)}+8A^{(0)}}\textrm{Im}\delta\tau}{(\textrm{Im}\tau^{(0)})^2} G_3^{(0)}\cdot\bar{G}_3^{(0)} +\frac{e^{2\Omega^{(0)}+8A^{(0)}}}{\textrm{Im}\tau^{(0)}} \left( \bar{G}_3^{(0)}\cdot W_3 - \bar{G}_3^{(0)}\cdot \chi_3 - \bar{\chi}_3\cdot G_3^{(0)} \right)\bigg]\,.
\end{align}
Due to the AISD property of $\bar G_3^\0$ and $W_3$, $\bar{G}_3^{(0)}\cdot W_3$ vanishes, and the definition of $\chi_3$ in \eqref{eq:chi3def} allows us to simplify this further as
\begin{align}
  \tilde{\nabla}^{\tilde{2}}\delta e^{-4A}-\tilde{\nabla}^{\tilde{m}}V_{m} + \frac{\bar{G}_3^{(0)}\cdot \tilde{d}\eta_2 + \tilde{d}\bar{\eta}_2\cdot G_3^{(0)}}{2\textrm{Im}\tau^{(0)}} = 0\,.  \label{eq:LapDelWarp}
\end{align}
With ISD background flux, this is the Hodge dual of \eqref{eq:F5Bianchi06} w.r.t. the background metric, and both are the first-order expansion of \eqref{eq:gkpbgwarpeqn}.

We can now determine the warp factor and Weyl factor through the use of Green's functions because \eqref{eq:LapDelWarp} takes the form of a Poisson equation of the form 
\begin{align}\label{eq:PoissonWDiv}
  \tilde{\nabla}^{\tilde{2}}\delta e^{-4A}=\tilde{\nabla}_mf^{\tilde{m}}\,,\quad f_{m}=V_{m}-\frac{1}{4\textrm{Im}\tau^{(0)}}\left(\eta_{np}\bar{G}^{(0)}{}_m{}^{\widetilde{np}}+\bar{\eta}_{np}G^{(0)}{}_m{}^{\widetilde{np}}\right)
\end{align}
after some simplification (because $G_3^\0$ is ISD and closed).
This can be solved in terms of the biscalar Green's function $\tilde{G}(y,Y)$ or equivalently in terms of the bivector Green's function $\tilde{G}_{\slashed{m} n}(y,Y)$ (see appendix \ref{app:BiscaBivec}) as
\begin{align}
  \delta e^{-4A}=\int d^6Y \sqrt{\tilde{g}_{(0)}(Y)}\t G(y,Y)\tilde{\nabla}_{\slashed{m}} f^{\tilde{\slashed{m}}}(Y)=\tilde{\nabla}_n\left[\int d^6Y \sqrt{\tilde{g}_{(0)}(Y)} \t G_{\slashed{m}}{}^{\tilde{n}}(y,Y) f^{\tilde{\slashed{m}}}(Y)\right]\,.
\end{align}
Here, slashed indices transform under diffeomorphisms at the point $Y$, as opposed to the point $y$. The above solution leads to vanishing of $\delta e^{-2\Omega}$, as follows. From the definition of $e^{-2\Omega}$ in \eqref{eq:defOmega}, for traceless metric fluctuations we have
\begin{align}
  \delta e^{-2\Omega}&=\frac{1}{\tilde{V}_{(0)}}\int d^6y \sqrt{\tilde{g}_{(0)}}\delta e^{-4A}\\
  &=\frac{1}{\tilde{V}_{(0)}}\int d^6y \int d^6Y \sqrt{\tilde{g}_{(0)}(y)}\sqrt{\tilde{g}_{(0)}(Y)} \tilde{\nabla}_n \t G_{\slashed{m}}{}^{\tilde{n}}(y,Y) f^{\tilde{\slashed{m}}}(Y) =0\,,
\end{align}
by interchanging the order of integration and subsequently owing to the fact that the internal manifold is closed (with some assumptions about smoothness of the Green's functions at the local sources).
The fact that a traceless deformation around $\tilde g_{mn}^{\0}$ does not change the overall warped volume is intuitive \cite{Douglas:2007tu}, but this is the first proof given in the literature to our knowledge.

The remaining SUGRA EOM are satisfied for constant $\delta\tau$.
In that case, the $G_3$ equation of motion acquires $[4,4]$ components at linear order. Because $G_3^\0$ is closed and ISD, many terms cancel automatically, leaving
\begin{align}
%  &4 e^{4\Omega^{(0)}}\hat{\epsilon}\wedge\left(\delta\Omega\tilde{d}(e^{4A^{(0)}}\tilde{\star}^{(0)}G_3^{(0)}) + \tilde{d}(e^{4A^{(0)}}\delta A\tilde{\star}^{(0)}G_3^{(0)})\right) \nn\\%
%  &+ 
e^{4\Omega^{(0)}}\hat{\epsilon}\wedge\tilde{d}e^{4A^{(0)}}\wedge\tilde{\star}^{(0)}(\chi_3-W_3) + e^{4\Omega^{(0)}+4A^{(0)}}\hat{\epsilon}\wedge\tilde{d}\tilde{\star}^{(0)}(\chi_3-W_3) 
%  &\qquad 
=ie^{4\Omega^{(0)}}\hat{\epsilon}\wedge\tilde{d}e^{4A^{(0)}}\wedge\chi_3 %+ 4ie^{4\Omega^{(0)}}\hat{\epsilon}\wedge\left(\delta\Omega\tilde{d}e^{4A^{(0)}}+\tilde{d}(e^{4A^{(0)}}\delta A)\right)\wedge G_3^{(0)}\,. 
\, .\label{eq:G3eom44Static}
\end{align}
The linear ISD condition \eqref{eq:ISDG3} implies that this equation is satisfied because $\chi_3$ is closed, just as we would expect for a GKP background.
Finally, for any constant $\delta\tau$, fluctuations of the $\tau$ equation of motion arise only through the term $G_3\wedge\star G_3$, given by
\begin{align}
  e^{4\Omega^{(0)}+4A^{(0)}}\hat{\epsilon}\wedge G_3^{(0)}\wedge\left(\tilde{\star}^{(0)}(\chi_3-W_3)-i\chi_3\right)\,,
\end{align}
which vanishes due to the condition \eqref{eq:ISDG3}.

\subsection{K\"ahler moduli}
\label{sec:GeoModKahler}

To understand geometric moduli of the CY, it is useful to work in complex coordinates. In these coordinates, 
K\"ahler moduli are fluctuations of the form $\delta_u\t g_{i\bar\jmath}=u\delta\tilde{g}_{i\bar{\jmath}}$ that preserve the Hermitian structure. 
Such deformations that keep the unwarped metric Ricci flat, i.e., solve \eqref{eq:metricmodcons}, are given by the components of a harmonic $(1,1)$-form $\omega_2$:\footnote{An $(i,j)$-form on the internal complex manifold refers to a differential form with $i$ holomorphic and $j$ antiholomorphic indices.}
$\delta g_{i\bar{\jmath}}=-i\omega_{i\bar{\jmath}}$ \cite{Candelas:1990pi}.
This deformation in the unwarped metric induces a shift in the K\"ahler form $\delta\tilde{J}_2=\omega_2$. 
Since we require traceless $\delta\t g_{mn}$, $\omega_2\cdot \t J_2^\0=0$. (This inner product is constant because the contraction of $\t J_2^\0$ with a harmonic form is harmonic.)

The expectation from \cite{Giddings:2001yu} is that K\"ahler deformations preserve the ISD condition, so all the K\"ahler moduli of the CY metric are still moduli of the full compactification.
In the conventions of \cite{Giddings:2001yu}, ISD 3-forms have $(0,3)$ and primitive $(2,1)$ components, where primitivity means that the wedge product with the K\"ahler form vanishes (i.e., $G_3\wedge\t J_2=0$).
As a result, the background 3-form flux $G_3^{(0)}$ has only $(0,3)$ and $(2,1)$ components and is primitive with respect to $\t J_2^\0$. In fact, because $G_3^\0$ is harmonic, it is automatically primitive; as we have seen, $G_3^\0\wedge\t J_2^\0$ is harmonic with respect to the background metric, so it must vanish because there are no harmonic 5-forms on a CY.
If the compactification contains a torus factor, there are nonvanishing harmonic 5-forms, so primitivity is an extra constraint. We discuss compactifications with torus factors further below, after the more general CY case.

From the definition \eqref{eq:W3def}, $W_3$ can have only have $(2,1)$ and $(0,3)$ components. 
Now from the condition \eqref{eq:ISDG3} we deduce that $\delta\tau=0$; otherwise, this would introduce $(3,0)$ and $(1,2)$ components to $W_3$ via $\bar{G}_3^{(0)}$.
Moreover, for traceless metric fluctuations, one can show that the $(0,3)$ component of $W_3$ vanishes, leaving a $(2,1)$ component that is non-primitive since $W_3$ is AISD. 
Therefore, \eqref{eq:ISDG3} is an equation for $\eta_2$ where $\chi_3=\tilde{d}\eta_2$ is a $(2,1)$ form.\footnote{since the Hodge dual of a $(2,1)$ form w.r.t. a K\"ahler metric gives another $(2,1)$ form.}

Rather than solving \eqref{eq:ISDG3}, we instead follow an argument by \cite{Martucci:2016pzt}. Since $\chi_3$ is $(2,1)$, $G_3$ has $(0,3)$ and $(2,1)$ components and must be primitive with respect to the deformed K\"ahler form
\begin{align}
  (G_3^{(0)}+u\chi_3)\wedge(\tilde{J}_2^{(0)}+u\delta\tilde{J}_2)=0\,
\Rightarrow  \chi_3\wedge\tilde{J}_2^{(0)}=-G_3^{(0)}\wedge\omega_2\,. \label{eq:PrimtivePertG3Lin}
\end{align}
It is sometimes assumed that $G_3^\0\wedge\omega_2$ vanishes because CY manifolds have no harmonic 5 forms, but the wedge product of harmonic forms is generically not harmonic (as explained in section \ref{s:harmonicproducts}).
$G_3^\0\wedge\omega_2$ is then exact because closed (due to Hodge decomposition theorem). 
Note that the background $G_3^\0$ is not primitive with respect to the deformed K\"ahler form, which is precisely because it is not harmonic with respect to the deformed metric (the harmonic representative of the cohomology class changes with the change in metric). The shift by $u\chi_3$ is precisely that needed to restore primitivity and harmonicity.

By the $\del\bar\del$ lemma, $\chi_3=2i\partial\bar{\partial}\Lambda_1$, where $\Lambda_1$ is a $(1,0)$ form.
Taking the background Hodge dual of \eqref{eq:PrimtivePertG3Lin}, using $\tilde{J}_2^{(0)}\propto\tilde{\star}^{(0)}(\tilde{J}_2^{(0)}\wedge \tilde{J}_2^{(0)})$, and assuming $\t\star^\0 \bar\del\t\star^\0 \Lambda_1=0$, we obtain \cite{Martucci:2016pzt}
\begin{align}
2i\left(\t\star^\0\del\t\star^\0\bar\del\right)\Lambda_1= \t\star^\0\left(G_3^\0\wedge\omega_2\right)\,\Rightarrow\quad
\triangle^\0\Lambda_1 = i\t\star^\0 \left(G_3^\0\wedge\omega_2\right)\, .
\end{align}
The operator on the left-hand-side is the Dolbeault Laplacian because $\t\star^\0\del\t\star^\0$ always vanishes on $(1,0)$ forms; this is half the Hodge-de Rham Laplacian $\triangle^\0$.
Since this Poisson equation is solvable in terms of bivector Green's functions, all K\"ahler moduli of the CY metric remain moduli of the GKP compactification.

Finally, we want to find $\eta_2$ such that $\t d\eta_2 = 2i\del\bar\del\Lambda_1$ and $\t d\t\star^\0\eta_2=0$. 
Taking $\eta_2=i(\bar\del-\del)\Lambda_1$ satisfies both requirements since the $\del$ and $\bar\del$ Laplacians are equal and $\bar\del\t\star^\0\Lambda_1=0$ implies $\bar\del\t\star^\0\bar\del\Lambda_1=0$ for a CY metric.

For a CY compactification with a formal metric, $G_3^\0\wedge\omega_2$ must vanish because it would be a harmonic 5-form, which does not exist (so $\Lambda_1$ and $\eta_2$ vanish).
The situation is somewhat different when the compactification contains a torus factor accounting for the formality of the flat metric on a torus.
The three cases of this type are $T^6/Z_2$ with O3-planes and $T^4\times T^2/Z_2$ or K3$\times T^2/Z_2$ with O7-planes.
For the O7 cases, the orientifold projection implies that $G_3^\0$ always has one leg along the $T^2/Z_2$ factor.
For the two torus compactifications, $G_3^\0\wedge\omega_2$ is harmonic but need not vanish because the first Betti number of the torus is nonzero; $G_3^\0\wedge\omega_2$ may be either exact or a nonvanishing harmonic form on K3$\times T^2/Z_2$. 
For nonvanishing harmonic $G_3^\0\wedge\omega_2$, it is impossible to solve \eqref{eq:PrimtivePertG3Lin} for exact $\chi_3$, and the K\"ahler modulus associated with $\omega_2$ is stabilized, as expected \cite{Kachru:2002he,Cicoli:2022vny}. We leave the dimensional reduction of these stabilized moduli to future work.
This implies that any massless K\"ahler modulus on a $T^6/Z_2$ or $T^4\times T^2/Z_2$ compactification has $\eta_2=0$.
The axionic partners of these stabilized moduli are fluctuations $\delta C'_4$ which can be removed by gauge transformations of the $A_2$ potential. Due to the nonvanishing first Betti number, there are vectors $A_{\mu m}$ in the 4D theory; the axions are Goldstone bosons for the gauge symmetry and are eaten via the St\"uckelberg mechanism \cite{Frey:2002hf,Frey:2003tf,Frey:2013bha}.

\subsection{Complex structure moduli}\label{sec:GeoModComplex}

Complex structure deformations of the background Ricci flat metric $\tilde{g}^{(0)}_{i\bar{\jmath}}$ are of the form $\bar u\delta^- \tilde{g}_{ij},u\delta^+\tilde{g}_{\bar{\imath}\bar{\jmath}}$, where $u,\bar u$ and $\delta^+\tilde{g}_{\bar{\imath}\bar{\jmath}}, \delta^- \tilde{g}_{ij}$ are complex conjugate pairs due to the reality of the line element (we use $\pm$ to indicate which variable pairs with $u,\bar u$ respectively). 
Moreover, the metric deformations can be respectively given in terms of a harmonic $(2,1)$ form $\xi_{\bar{\imath}kl}$ and the conjugate $(1,2)$ from $\bar{\xi}_{i\bar{k}\bar{l}}$ as \cite{Candelas:1990pi}
\begin{align}
  \delta^- \tilde{g}_{ij}=\bar{\xi}_{(i}{}^{kl}\tilde\Omega^\0_{j)kl}\,,\quad \delta^+\tilde{g}_{\bar{\imath}\bar{\jmath}}=\xi_{(\bar{\imath}}{}^{\bar{k}\bar{l}}\bar{\tilde\Omega}^\0_{\bar{\jmath})\bar{k}\bar{l}}\,, \label{eq:CompStrDefMet}
\end{align}
where $\t\Omega_3^\0$ is the holomorphic harmonic $(3,0)$-form w.r.t. the background metric (and $\bar{\t\Omega}_3^\0$ is its conjugate).\footnote{As usual, in this section, indices are raised by background inverse metric $\tilde{g}^{i\bar{\jmath}}_{(0)}$.} 
The harmonic $(2,1)$ forms are all primitive with respect to $\t J_2^\0$ following the same reasoning as given for $G_3^\0$ above (modulo the same caveats regarding torus factors).
We note that the (2,1) forms $\xi_3$ (and $\xi_G$ below) are harmonic with respect to the background unwarped metric, but we do not add tildes or superscript $\0$ to avoid making the notation too cumbersome.

As indicated, a geometric complex structure modulus $u$ is naturally a complex variable. While we can work with linear combinations of $\delta^-\t g_{ij}$ and $\delta^+\t g_{\bar\imath\bar\jmath}$ corresponding to the real and imaginary parts of $u$, we will typically consider complex variables.
Therefore, $\delta^-\t g_{ij}$ and $\delta^+\t g_{\bar\imath\bar\jmath}$ ($u$ and $\bar u$) can be treated as independent in linear perturbation theory.
However, because the line element must remain real, we solve the linearized ISD condition for both conjugate deformations simultaneously.
The combined linear ISD condition, generalizing \eqref{eq:ISDG3}, is
\begin{align}
  \delta_u W_3 =\tilde{d}\delta_u A_2+i\tilde{\star}^{(0)}\t d\delta_u A_2+\frac{i\delta_u\tau}{\textrm{Im}\tau^{(0)}}\bar{G}_3^{(0)}\,, \label{eq:ISDG3full}
\end{align}
where we use
\be
  &\delta_u W_3 \equiv uW^+_3+\bar u W_3^-\,, \quad W^\pm_{mnp} \equiv 3\delta^\pm \tilde{g}_{l[m}G^{(0)}{}^{\tilde{l}}{}_{np]}\,, \\
  &\delta_u A_2 \equiv u\eta^+_2+\bar u\eta^-_2\,, \quad \delta_u \tau \equiv u\delta^+\tau+\bar u\delta^-\tau\,. \label{eq:pmDefs}
\ee

The expectation of \cite{Giddings:2001yu} was that complex structure or axiodilaton deformations violate the ISD condition and stabilize the moduli by changing the Hodge type of $\delta\t\star G_3^\0$ relative to $G_3^\0$ (in our terms, \eqref{eq:ISDG3full} is not solvable).
However, for a given background flux, some complex structure moduli, perhaps paired with a specific shift in the axiodilaton, can satisfy \eqref{eq:ISDG3full} and remain moduli of the full GKP compactification. 
We will call these complex structure flat directions since they do not include all of the geometric complex structure moduli space.
For large numbers ($h^{2,1}$) of complex structure moduli, the tadpole conjecture suggests that there may be many such flat directions.
If $u,\xi_3$ and $\bar u,\bar\xi_3$ satisfy separate relations, the flat direction is complex, and we continue to treat $u$ and $\bar u$ as independent.
On the other hand, if $u,\bar u$ are related by \eqref{eq:ISDG3full}, the flat direction is real, so we can work with a single real deformation from the beginning or take the appropriate linear combination of the complex deformations.
Here we will give some conditions on the complex structure flat directions assuming that they obey the linearized ISD condition \eqref{eq:ISDG3full} and when they are real or complex.

We write background harmonic flux $G_3^{(0)}$ as
\begin{align}
  G_3^{(0)}=\lambda \bar{\t\Omega}_3^\0+\xi_G\,,
\end{align}
where $\lambda$ is a proportionality constant and $\xi_G$ is a (primitive) harmonic $(2,1)$ form. 
For the holomorphic complex structure deformations $u$ ($\delta^+ \tilde{g}_{\bar{\imath}\bar{\jmath}}$), $W^+_3$ has a $(1,2)$ component given by
\begin{align}\label{eq:CompStruc2}
  W^+_{i\bar{\jmath}\bar{k}}=3\delta^+\tilde{g}_{\bar{l}[\bar{\jmath}}\xi_G{}^{\bar{l}}{}_{\bar{k}i]} = \frac{3}{2} \left(\xi{}^{l\bar{a}\bar{b}}\bar{\t\Omega}^\0_{\bar{a}\bar{b}[\bar{\jmath}}\xi_G{}_{\bar{k}i]l} + \xi_{[\bar{\jmath}}{}^{\bar{a}\bar{b}}\xi_G{}{}_{\bar{k}i]l}\bar{\t\Omega}^{\0 l}{}_{\bar{a}\bar{b}}\right)\,.
\end{align}
Because the harmonic $(3,0)$ form is similar to the volume form for holomorphic indices (in fact, $\t\epsilon\propto\t\Omega\wedge\bar{\t\Omega}$), it satisfies the identity
\be\label{eq:OmegaIdent}
\t\Omega^{mnr}\bar{\t\Omega}_{pqr} = \frac{1}{2} \delta^{m}_{[p}\delta^n_{q]}-\frac 12 \t J_{[p}{}^m \t J_{q]}{}^n+i\delta^{[m}_{[p}\t J_{q]}{}^{n]}\,\Rightarrow\quad 
\t \Omega^{\bar\imath\bar\jmath\bar k}\bar{\t \Omega}_{\bar a\bar b\bar k} = 2\delta^{\bar\imath}_{[\bar a}\delta^{\bar\jmath}_{\bar b]}\,,
\ee
so we can rearrange \eqref{eq:CompStruc2} into a more instructive form
\be\label{eq:CompW2}
W^+_{i\bar\jmath\bar k} =& 2\left(\xi_G{}_{i[\bar\jmath}{}^{\bar a} \xi^{\bar b\bar c}{}_{\bar k]}\right) \bar{\t\Omega}^\0_{\bar a\bar b\bar c}\,.
\ee
In general, this may contain harmonic, exact, and coexact components.

For antiholomorphic deformations $\bar u\delta^- \tilde{g}_{ij}$, $W^-_3$ acquires $(3,0)$ and $(1,2)$ components, given by
\begin{align}\label{eq:CompStruc1}
  &W^-_{ijk}=3\delta^-\tilde{g}{}^{\bar{l}}{}_{[i}\xi_G{}_{jk]\bar{l}} = \frac{3}{2} \left(\bar{\xi}{}^{\bar{l}ab}\t\Omega^\0_{ab[i}\xi_G{}_{jk]\bar{l}}+\bar{\xi}_{[i}{}^{ab}\xi_G{}_{jk]\bar{l}}\t\Omega^{\0\bar{l}}{}_{ab}\right)\,, \nn\\
  &W^-_{i\bar{\jmath}\bar{k}}=3\lambda \delta^-\tilde{g}{}^{\bar{l}}{}_{[i}\bar{\t\Omega}^\0_{\bar{\jmath}\bar{k}]\bar{l}} = \frac{3\lambda}{2} \left(\bar{\xi}{}^{\bar{l}ab}\t\Omega^\0_{ab[i}\bar{\t\Omega}^{\0}_{\bar{\jmath}\bar{k}]\bar{l}}+\bar{\xi}_{[i}{}^{ab}\bar{\t\Omega}^{\0}_{\bar{\jmath}\bar{k}]\bar{l}}\t\Omega^{\0\bar{l}}{}_{ab}\right)\,.
\end{align}
With the same simplifications as above, these become
\be\label{eq:CompW1}
W^-_{ijk} =& 2\t\Omega_{ijk}^\0 \left(\xi_G\cdot\bar\xi_3\right)  \, ,\quad
W^-_{i\bar\jmath\bar k} =& 2\lambda \bar\xi_{i\bar\jmath\bar k}\, .
\ee
The $(3,0)$ component may or may not be harmonic, depending on the $y$ dependence of the inner product of $\xi_3,\xi_G$; it is always harmonic for a formal metric where the inner product is constant.
The $(1,2)$ component is always harmonic. It is interesting to note that this $(1,2)$ component appears when the background flux has a $(0,3)$ component, which breaks SUSY.

We can now make some comments about the form of flat directions among the axiodilaton and geometric complex structure moduli by comparison to the linearized ISD condition \eqref{eq:ISDG3full}.
Recall that the total metric deformation must be real, so we must include both $u,\bar u$ when solving this equation.
\begin{itemize}
\item Any exact or co-exact component of $\delta_u W_3$ can always be compensated by $\delta_u A_2$, so the harmonic components in the Hodge decomposition determine the flat directions. It is possible to determine the harmonic components by wedging $\delta_u W_3$ with $\bar{\t\Omega}_3^\0$ or a basis $\xi^a_3$ of harmonic $(2,1)$ forms and integrating over the compact manifold.

\item If the background preserves SUSY, $\lambda=0$. In this case, the $(3,0)$ component of \eqref{eq:ISDG3full} depends on the $\bar\xi_3$ or $\bar u$ (antiholomorphic) moduli only, while the $(1,2)$ components depend on $\xi_3$ (the holomorphic $u$ moduli), and $\delta_u\tau=u\delta^+\tau$. That is, the holomorphic and antiholomorphic variables do not mix, which is consistent with SUSY. In other words, it is consistent to treat $u$ and $\bar u$ as independent flat directions with independent compensators $\eta^\pm_2$ while carrying out the dimensional reduction at this order.

\item On the other hand, $\lambda\neq 0$ breaks SUSY, and both holomorphic and antiholomorphic deformations appear in the same component of \eqref{eq:ISDG3full} (the ISD condition may only be solvable for a linear combination of holomorphic and antiholomorphic moduli). In other words, the flat directions may be the real or imaginary part only of a geometric complex structure modulus and not a complex variable. Both $\delta_u A_2$ and $\delta_u\tau$ contain $u$ and $\bar u$, but it is convenient to write the flat direction in terms of a single real parameter $u$ (with a single $\eta_2$ and $\delta\tau$) because the holomorphic and antiholomorphic variables are related. The SUSY-breaking example of \cite{Frey:2002qc} has flat directions of this type for a torus compactification, and \cite{Hebecker:2020ejb,Cicoli:2022vny} discuss real flat directions for CY compactifications.

\item There is a flat direction if and only if any harmonic component of $\delta_u W_3$ is proportional to $\bar G_3^\0$, and the proportionality constant determines $\delta_u\tau$. That is, two conditions hold: $2\lambda\xi_3$ plus any harmonic contribution from \eqref{eq:CompW2} is proportional to $\xi_G$, and any harmonic $(3,0)$ component of $\delta_u W_3$ is the same proportionality constant times $\lambda\tilde\Omega_3$.

\item Conversely, a shift $\delta_u \tau$ in the axiodilaton must be accompanied by some complex structure deformation; the axiodilaton direction can never be a flat direction on its own.

\item In the 4D effective theory, the F-flatness conditions $D_\tau W =D_aW=0$ of the SUGRA yield the ISD condition, where $W$ is the Gukov-Vafa-Witten superpotential and $D_\tau,D_a$ are the K\"ahler covariant derivatives with respect to the axiodilaton and a basis of complex structure deformations. Since we consider fluctuations around a fixed GKP background, the harmonic components of \eqref{eq:ISDG3full} should correspond to the variation of $D_\tau W$ and $D_a W$ along $\xi_3,\delta\tau$. We will discuss this point in more detail in section \ref{s:kineticcomplex}.
\end{itemize}
In summary, the harmonic parts of $\delta_u W_3$ cancel with the $\delta_u \tau$ term of \eqref{eq:ISDG3full} for a flat direction, leaving only the exact and co-exact parts. This somewhat complicated form leads us to believe that there is no general solution for $\eta^\pm_2$, but they can be found by imposing the Poisson equations
\begin{align}
  \tilde{d}\tilde{\star}^{(0)}\tilde{d}\eta^{\pm}_2=-i\tilde{d}W^{\pm}_3\, , \label{eq:Eta2pmEqn}
\end{align}
the analogues of \eqref{eq:Eta2EqnDirect}.
Because the Laplacian of the background K\"ahler metric preserves the number of holomorphic/antiholomorphic indices, this Poisson equation implies that $\eta^+_2$ has $(1,1)$ and $(0,2)$ components
while $\eta_2^-$ has only a $(2,0)$ component ($\t dW_3^-$ is $(3,1)$ because the $(1,2)$ component of $W_3^-$ is harmonic).

When the compact metric $\tilde{g}_{mn}$ is formal, $W^-_3$ in \eqref{eq:CompW1} is harmonic; it is natural to speculate that $W^+_3$ as in \eqref{eq:CompW2} is also harmonic in this case, so $\eta_2$ for these moduli vanishes. However, it is not clear whether that particular product of harmonic forms is necessarily harmonic in a formal metric.

We can say more for $T^6/Z_2$ or $T^4\times T^2/Z_2$ torus compactifications. (The K3$\times T^2/Z_2$ case fits in the general framework above.) 
On the torus, harmonic forms are exactly those forms with constant components, so $W_3$ as given in (\ref{eq:CompW2},\ref{eq:CompW1}) is always harmonic. Therefore, $W_3$ must either vanish or be proportional to $\bar G_3^\0$ for any complex structure flat direction, and $\eta_2^\pm=0$.
We consider some examples on $T^6/Z_2$ compactifications as a testing ground for these conditions due to the simple nature of the harmonic forms on $T^6$. 
Our examples come from \cite{Cicoli:2022vny}, which analyzed supersymmetric background values of $G_3^\0$ (subject to the tadpole constraint) and classified them by the number of complex structure flat directions as given by the 4D effective superpotential. 
%In addition, maintaining the D3 tadpole bound the flux quanta (in other words, background values of $G_3^{(0)}$) were classified in \cite{Cicoli:2022vny} based on the number of complex structure flat directions the respective flux vacua admit. (Their classification directly analyzed the 4D EFT superpotential.) 
In appendix \ref{app:CompStrFlat}, we examine three such flux backgrounds and for each of them determine which complex structure deformations (if any) satisfy \eqref{eq:ISDG3full}. Our results are in complete agreement with those found in \cite{Cicoli:2022vny}. Furthermore, the explicit examples illustrate how to select a harmonic $(2,1)$ form $\xi$ (corresponding to a flat direction) to use in $ \eqref{eq:CompStrDefMet}$. 
%Below, we highlight the results for one such example; for details, see appendix \ref{app:CompStrFlat}.

Here we highlight features of one example of a GKP background that admits one complex structure flat direction; we give details in appendix \ref{app:CompStrFlat}. 
Following \cite{Cicoli:2022vny}, the background 3-form for this example is a harmonic $(2,1)$ form, given in complex coordinates by
\begin{align}
  G^{(0)}_3 = \frac{1}{2i \, \text{Im} \, \tau^{(0)}} \left( 2 dz^1 \wedge dz^2 \wedge d\bar{z}^3 - 2 dz^1 \wedge dz^3 \wedge d\bar{z}^2 - 4 dz^2 \wedge dz^3 \wedge d\bar{z}^1 \right) \,.
\end{align}
Here, the background holomorphic 1-forms are given by $dz^i = dy^i + \tau^{(0)}_i \, dy^{i+3}$ for $i = 1, 2, 3$, where the background values of the complex structure moduli $\tau^{(0)}_i$ are equal to the background value of the axiodilaton $\tau^{(0)}$, and $(y^1, \dots, y^6)$ denote the real periodic right-handed coordinates on $T^6$. When we allow deformations $\tau_i^{(0)} \to \tau_i^{(0)} + u\delta\tau_i$, the resulting metric deformations, contracted with $G_3^{(0)}$, generate a harmonic $W_3$ form with non-trivial constant $(3,0)$ and $(1,2)$ components.\footnote{Note that each complex structure deformation has a single complex parameter $\delta\tau_i$ corresponding to a $(2,1)$ form $\xi^i_3$ in the metric deformation $\delta^+\t g_{\bar\imath\bar\jmath}$. The $\delta\bar\tau_i$ parameters appear in $\delta^-\t g_{ij}$.} 
Then, including the axiodilaton fluctuation $\delta_u\tau=u\delta^+\tau$, \eqref{eq:ISDG3full} reduces to four linear equations
\begin{align}
  &\delta\bar\tau_2 +\delta\bar\tau_3=2\delta\bar \tau_1\,, \quad \delta \tau_1+ \delta^+ \tau=2 \delta \tau_2\,, \nn\\
  &\delta \tau_2+ \delta \tau_3=2 \delta^+ \tau\,, \quad\delta \tau_1+\delta^+ \tau=2\delta \tau_3\,.
\end{align}
This condition yields a flat direction with all complex structure deformations $\delta\tau_i=\delta^+\tau$, the axiodilaton deformation; since it is consistent to set $\delta^-\tau=0$, $u$ and $\bar u$ do not mix, and the flat direction is complex.
For this flat direction, the harmonic $(2,1)$ form $\xi_3$ can be taken as a linear combination given by
\begin{align}
  \xi_3 \propto dz^1 \wedge dz^2 \wedge d\bar{z}^3 + dz^2 \wedge dz^3 \wedge d\bar{z}^1 + dz^3 \wedge dz^1 \wedge d\bar{z}^2
\end{align}
which can be used in \eqref{eq:CompStrDefMet}. In appendix \ref{app:CompStrFlat}, we also discuss one example with two complex structure flat directions and another with none.

\section{Linearized equations of motion around warped background}\label{s:linearization}

The moduli space metric appears in the kinetic action of the effective field theory for the moduli. Therefore, we must promote flucutations of the GKP background to $x$-dependent functions $u(x)$.
In this section, we derive the constraint equations and EOM operators based on an ansatz for linear fluctuations of various field around GKP background. We will also see how our ansatz solves the constraints, which are required for gauge invariance.
Because we study motion on the moduli space of the GKP compactifications, we must impose the relations between fields that we derived in the previous section. Since we work at linear order, we can consider a single modulus, so all 10D field fluctuations will have the same spacetime dependence $u(x)$.

\subsection{Ansatz for linear fluctuations}

We consider the following ansatz for the 10D metric for traceless geometric moduli of $\t g_{mn}$:
\begin{align}
  ds_{10}^2=e^{2\Omega+2A(x,y)}\hat{\eta}_{\mu\nu}dx^{\mu}dx^{\nu}+2e^{2\Omega+2A(x,y)}\partial_\mu u(x)B_m(y)dx^{\mu}dy^m+e^{-2A(x,y)}\tilde{g}_{mn}(x,y)dy^m dy^n\, , \label{eq:metricansatz}
\end{align}
where up to linear order the coordinate dependencies are
\begin{align}
\tilde{g}_{mn}(x,y)=\tilde{g}^{(0)}_{mn}(y)+u(x)\delta\tilde{g}_{mn}(y)\,,\quad \delta\tilde{g}^{kl}\equiv-\tilde{g}^{kp}_{(0)}\tilde{g}^{lq}_{(0)}\delta\tilde{g}_{pq}\,,
  \quad A(x,y)=A^{(0)}(y)+u(x)\delta A(y)\,.
\end{align}
We have seen in section \ref{s:staticmod} that the Weyl factor $\Omega$ is unaffected by the fluctuation, i.e. $\delta\Omega=0$.

We also consider the following ansatz for form fluxes and axiodilaton:
\begin{align}
  &\tilde{F}_5=\tilde{\star}\tilde{d}e^{-4A}-e^{2\Omega}\hat{d}u\wedge\tilde{\star}\tilde{d}B_1+\left\{e^{4\Omega}\hat{\epsilon}\wedge \tilde{d}e^{4A}+e^{4\Omega}\hat{\star}\hat{d}u\wedge B_1\wedge \tilde{d}e^{4A}-e^{4\Omega+4A}\hat{\star}\hat{d}u\wedge\tilde{d}B_1\right\}\,, \nn\\
  &G_3=G_3^{(0)}(y)+\hat{d}u(x)\wedge \eta_2(y)+u(x)\chi_3(y)\,,\quad \tau(x,y)=\tau^{(0)}+u(x)\delta\tau\,, \label{eq:formsansatz}
%  &B_1\equiv B_m(y)dy^m\,,\quad \hat{\epsilon}\equiv\frac{1}{4!}\hat{\epsilon}_{\mu\nu\rho\sigma}dx^\mu\wedge dx^\nu\wedge dx^\rho\wedge dx^\sigma\,. \nn
\end{align}
where $B_1$ is the same 1-form as appears in the metric \eqref{eq:metricansatz}.
This ansatz for $\tilde{F}_5$ maintains self-duality up to linear order; we have stated the independent magnetic ($[0,5]$ and $[1,4]$) components of the 5-form first and the additional components required for self-duality in brackets for convenience. In the 3-form, $\eta_2,\chi_3$ are differential forms on the internal manifold related by \eqref{eq:chi3def}. $\delta\tau$ is a constant as determined in section \ref{sec:GeoModKahler} or \ref{sec:GeoModComplex}. 

% In above ansatzes \eqref{eq:metricansatz} and \eqref{eq:formsansatz}, all the $x$-dependencies are expressed in terms of geometric modulus $u(x)$ which we consider as infinitesimal. The coefficients in the linear fluctuations namely $\delta\tilde{g}_{mn}, B_m, \delta A, \eta_2, \chi_3$ are $y$-dependent while $\delta\Omega, \delta\tau$ are some constants.  \rma{Comment that we could have allowed additional terms in $\delta A_2,\tilde{F}_5$, and a $y$-dependence $\delta\tau(y)$ at the onset, but ended up setting them to zero consistently.}

The ansatz for $G_3$ involves allowing a linear fluctuation of the form $\delta_u A_2 = u(x)\eta_2(y)$ in \eqref{eq:G3fluctuation}. The $x$-dependence of $u$ generates the term involving $\hat{d}u$, while $\chi_3$ remains defined as in \eqref{eq:chi3def}. It is also straightforward to check that above ansatz for $G_3$ satisfies the Bianchi identity \eqref{eq:bianchi2}. 

% Clearly, the ansatz for $\tilde{F}_5$ consists of two components which are related by the Hodge duality, namely $\tilde{\star}\tilde{d}e^{-4A}-e^{2\Omega}\hat{d}u\wedge\tilde{\star}\tilde{d}B_1$ and the remainder. In terms of the fluctuations in the form fields $C_4, C_2, B_2$, the fluctuation in $\tilde{F}_5$ can be expressed as\footnote{$\delta_u C_4^\prime\equiv \delta_u C_4-C_2^{(0)}\wedge\delta_u B_2/2+B_2^{(0)}\wedge\delta_u C_2/2$, as in section \ref{sec:review.dim.reduce}.}

Due to the self-duality of $\tilde{F}_5$ (up to linear order), not all components of the fluctuations of $C_4$ or $\t F_5$ are independent, as described in section \ref{sec:review.dim.reduce} above.
We consider the ``magnetic'' $[0,5]$ and $[1,4]$ components to be independent; the 5-form fluctuations in these components are (recall \eqref{eq:fluctF05})
\be\label{eq:F5mag1}
\delta_u\tilde{F}_{5}=-u\tilde{\star}^{(0)}V_1+u\tilde{\star}^{(0)}\tilde{d}\delta e^{-4A} -e^{2\Omega}\hat{d}u\wedge\tilde{\star}\tilde{d}B_1\, .
\ee
In terms of the SUGRA fields $C_4, C_2, B_2$, the fluctuation in the magnetic components of $\t F_5$ should be expressed as
\begin{align}
  \delta_u\tilde{F}_5&=d\delta_u C_4^\prime+\frac{i}{2\textrm{Im}\tau^{(0)}}\left(\delta_uA_2\wedge\bar{G}_3^{(0)}-\delta_u\bar{A}_2\wedge G_3^{(0)}\right) 
 % &= d\delta_u C_4^\prime+\frac{iu}{2\textrm{Im}\tau^{(0)}}\left(\eta_2\wedge\bar{G}_3^{(0)}-\bar{\eta}_2\wedge G_3^{(0)}\right)
 \,. \label{eq:FlucDefF5}
\end{align}
%
 %This will be revisited in section \ref{sec:ModuliActionMetric}, where we derive the kinetic action and retain only the magnetic components of $\delta_u C_4^\prime$, given by 
We will see below that \eqref{eq:F5mag1} takes the form \eqref{eq:FlucDefF5} with $\delta_u C'_4=-e^{2\Omega^{(0)}} u \tilde{\star} \tilde{d}B_1$.

\subsection{Constraints and dynamical EOM}
\label{sec:Constr.dynEOM}

Substituting the ansatzae \eqref{eq:metricansatz} and \eqref{eq:formsansatz} into the SUGRA equations of motion and Bianchi identities \eqref{eq:eomforms2}–\eqref{eq:eomeinstein} yields a variety of constraints and EOM operators at linear order. 
See appendix \ref{app:FlucTermsEOMBian} for details of the linearization procedure. 
Recall that restricting to (unstabilized) moduli of the GKP compactification means that terms proportional to $u$ (and not its derivatives) must vanish separately; these are the constraints that we discussed in section \ref{s:staticmod} above.

\paragraph{$\tilde{F}_5$ Bianchi identity:} Because we will retain only the magnetic components of the 4-form when we find the effective action, it is useful to consider the 5-form Bianchi identity and EOM as separate equations including only the components of $\t F_5$ outside of the curly brackets in \eqref{eq:formsansatz}.

The $\tilde{F}_5$- Bianchi identity leads to two constraints, as follows. First, the $[0,6]$ components (which are proportional to $u$) of the Bianchi identity at linear order give the constraint \eqref{eq:F5Bianchi06}.
Second, the $[1,5]$ components (which are wedged with $\hat{d}u$) at linear order give another constraint
\begin{align}
  \tilde{\star}^{(0)}\tilde{d}\delta e^{-4A} -\tilde{\star}^{(0)}V_1 +e^{2\Omega^{(0)}}\tilde{d}\tilde{\star}^{(0)}\tilde{d}B_1 -\frac{i}{2\textrm{Im}\tau^{(0)}}\left(\eta_2\wedge\bar{G}_3^{(0)}-G_3^{(0)}\wedge\bar{\eta}_2\right)=0\,. \label{eq:F5Bianchi15}
\end{align}

The linear fluctuation in $\tilde{F}_5$ \eqref{eq:F5mag1} that results from our ansatz can be brought to the following form using the constraint \eqref{eq:F5Bianchi15}:
\begin{align}
  \delta_u \tilde{F}_5=&-d\left(e^{2\Omega^{(0)}}u\tilde{\star}^{(0)}\tilde{d}B_1\right)+\frac{iu}{2\textrm{Im}\tau^{(0)}}\left(\eta_2\wedge\bar{G}_3^{(0)}-\bar{\eta}_2\wedge G_3^{(0)}\right)+\cdots\,.
\end{align}
Comparing this expression with \eqref{eq:FlucDefF5}, we can read off the magnetic components of $\delta C_4^\prime$ as
\begin{align}
  \delta_u C_4^\prime=-e^{2\Omega^{(0)}} u \tilde{\star} \tilde{d}B_1\,.
\end{align}
This shows that the ansatz \eqref{eq:formsansatz} for $\t F_5$ indeed takes the form that we expect from general principles and is a nontrivial consistency check.

\paragraph{$\t F_5$ equation of motion:}
The equation of motion $d\star\t F_5=\cdots$ has only $[4,2]$ components, which lead to a dynamical equation of motion 
\begin{align}
  e^{4\Omega^{(0)}}\hat{d}\hat{\star}\hat{d}u\wedge B_1\wedge \tilde{d}e^{4A^{(0)}(y)}-e^{4\Omega^{(0)}+4A^{(0)}}\hat{d}\hat{\star}\hat{d}u\wedge\tilde{d}B_1=0\,,
\end{align}
at linear order.
\emph{We do not require the fluctuation to satisfy this EOM since we work offshell.} 
However, we identify the linear fluctuation in the EOM operator $E_6$ as
\begin{align}
  \delta_u E_6=-e^{4\Omega^{(0)}}\hat{d}\hat{\star}\hat{d}u\wedge\tilde{d}\left(e^{4A^{(0)}}B_1\right)\,.
\end{align}

\paragraph{Einstein field equations:} Now we discuss the Einstein field equations. Equating the $\mu\nu$-components on both sides of the equations at linear order, we get
%
% \begin{align}
%   &\frac{u}{2}e^{2\Omega^{(0)}+8A^{(0)}}\left[\tilde{\nabla}^{\tilde{2}}\delta e^{-4A}-\tilde{\nabla}^{\tilde{m}}V_{m} + \frac{\bar{G}_3^{(0)}\cdot \tilde{d}\eta_2 + \tilde{d}\bar{\eta}_2\cdot G_3^{(0)}}{2\textrm{Im}\tau^{(0)}}\right]\hat{\eta}_{\mu\nu} \nn\\ &\qquad + 2\left(\delta\Omega-2\delta A-\frac{1}{2}e^{2\Omega^{(0)}+4A^{(0)}}\tilde{\nabla}^{\tilde{m}}B_m\right)\left(\partial^{\hat{2}}u\ \hat{\eta}_{\mu\nu}-\partial_{\mu}\partial_{\nu}u\right) = 0\,.
% \end{align}
% %
\begin{align}
  \left(-2\delta A-\frac{1}{2}e^{2\Omega^{(0)}+4A^{(0)}}\tilde{\nabla}^{\tilde{m}}B_m\right)\left(\partial^{\hat{2}}u\ \hat{\eta}_{\mu\nu}-\partial_{\mu}\partial_{\nu}u\right) = 0 \label{eq:munuEinFieldEqn}
\end{align}
since the term proportional to $u$ vanishes owing to \eqref{eq:LapDelWarp}.

Now, equating the $\mu m$-components on both sides of the equations at linear order, we get, up to an overall factor of $\partial_\mu u$:
\begin{align}
  &-\frac{1}{2}e^{4A^{(0)}}\partial_{m}\delta e^{-4A} + \frac{1}{2} e^{4A^{(0)}}V_{m} + \frac{1}{2}e^{2\Omega^{(0)}+4A^{(0)}}\tilde{\nabla}^{\tilde{n}}(\tilde{d}B_1)_{mn} \nn\\
  &\qquad\qquad -2e^{2\Omega^{(0)}+4A^{(0)}}\left(\tilde{\nabla}^{\tilde{2}}A^{(0)}-4\partial_{n}A^{(0)}\partial^{\tilde{n}}A^{(0)}\right)B_{m} +\frac{1}{2}\tilde{\nabla}^{\tilde{n}}\delta\tilde{g}_{mn} \nn\\
  &\qquad =-\frac{1}{4} B_m \frac{e^{2\Omega^{(0)}+8A^{(0)}}}{\textrm{Im}\tau^{(0)}}G_3^{(0)}\cdot\bar{G}_3^{(0)} +\frac{ie^{4A^{(0)}}}{4\textrm{Im}\tau^{(0)}} \Big[\tilde{\star}^{(0)}(\eta_2\wedge\bar{G}_3^{(0)}-\bar{\eta}_2\wedge G_3^{(0)})\Big]_m\,.
\end{align}
Using the background equation of motion for the warp factor $A^{(0)}$ (the Hodge dual of \eqref{eq:gkpbgwarpeqn}), we can simplify the above equation to
% \footnote{Taking the Hodge dual of \eqref{eq:gkpbgwarpeqn} w.r.t. $\tilde{g}^{(0)}_{mn}$ and utilizing the ISD property of the background 3-form $G_3^{(0)}$, one obtains $\tilde{\nabla}^{\tilde{2}}A^{(0)}-4\partial_nA^{(0)}\partial^{\tilde{n}}A^{(0)}=e^{4A^{(0)}}G_3^{(0)}\cdot\bar{G}_3^{(0)}/8\textrm{Im}\tau^{(0)}$.}
%
\begin{align}
  &-\frac{1}{2}e^{4A^{(0)}}\partial_{m}\delta e^{-4A} + \frac{1}{2} e^{4A^{(0)}}V_{m} + \frac{1}{2}e^{2\Omega^{(0)}+4A^{(0)}}\tilde{\nabla}^{\tilde{n}}(\tilde{d}B_1)_{mn} +\frac{1}{2}\tilde{\nabla}^{\tilde{n}}\delta\tilde{g}_{mn} \nn\\
  &\qquad\qquad =\frac{ie^{4A^{(0)}}}{4\textrm{Im}\tau^{(0)}} \Big[\tilde{\star}^{(0)}(\eta_2\wedge\bar{G}_3^{(0)}-\bar{\eta}_2\wedge G_3^{(0)})\Big]_m\,. \label{eq:mnEinFieldEqn}
\end{align}
Next, the $mn$-components at linear order give a dynamical equation of motion
\begin{align}
  &-\frac{1}{2}e^{-2\Omega^{(0)}-4A^{(0)}}\partial^{\hat{2}}u\delta\tilde{g}_{mn} + e^{-2\Omega^{(0)}-4A^{(0)}}\partial^{\hat{2}}u\left(-2\delta A\right)\tilde{g}_{mn} \nn\\
  &\quad+ \partial^{\hat{2}}u\tilde{\nabla}_{(m}B_{n)} + 4\partial^{\hat{2}}u\partial_{(
m}A^{(0)}B_{n)} - \partial^{\hat{2}}u(\tilde{\nabla}^{\tilde{p}}B_p)\tilde{g}^{(0)}_{mn} - 2\tilde{g}^{(0)}_{mn}\partial^{\tilde{p}}A^{(0)}B_{p}\partial^{\hat{2}}u =0\,,
\end{align}
\emph{which we do not require to satisfy offshell.} Instead, we identify the linear fluctuation in the EOM operator $E_{mn}$ as
\begin{align}
  \delta_u E_{mn}=&\partial^{\hat{2}}u\Big\{-\frac{1}{2}e^{-2\Omega^{(0)}-4A^{(0)}}\delta\tilde{g}_{mn} +\big(\frac{1}{2}e^{-2\Omega^{(0)}}\delta e^{-4A} \big)\tilde{g}_{mn} \nn\\
  &\quad+ \tilde{\nabla}_{(m}B_{n)} + 4\partial_{(
m}A^{(0)}B_{n)} - (\tilde{\nabla}^{\tilde{p}}B_p)\tilde{g}^{(0)}_{mn} - 2\tilde{g}^{(0)}_{mn}\partial^{\tilde{p}}A^{(0)}B_{p}\Big\}\,.
\end{align}

\paragraph{$G_3$ equation of motion:} Now we deal with the $G_3$ equation of motion. Equating the $[3,5]$ components on both sides of the equation at linear order leads to
\begin{align}
  &e^{4\Omega^{(0)}}\hat{\star}\hat{d}u\wedge\tilde{d}(e^{4A^{(0)}}B_1\wedge\tilde{\star}^{(0)}G_3^{(0)}) + e^{2\Omega^{(0)}}\hat{\star}\hat{d}u\wedge\tilde{d}\tilde{\star}^{(0)}\eta_2 \nn\\
  &\qquad\qquad =ie^{4\Omega^{(0)}}\hat{\star}\hat{d}u\wedge\tilde{d}e^{4A^{(0)}}\wedge B_1\wedge G_3^{(0)} + ie^{4\Omega^{(0)}+4A^{(0)}}\hat{\star}\hat{d}u\wedge\tilde{d}B_1\wedge G_3^{(0)}\,.
\end{align}
Using the ISD property and the closedness of the background 3-form flux $G_3^{(0)}$, the above equation becomes
\begin{align}
  &e^{2\Omega^{(0)}}\hat{\star}\hat{d}u\wedge\tilde{d}\tilde{\star}^{(0)}\eta_2=0\,, \label{eq:G3eom35}
\end{align}
so we see that choosing $\eta_2$ co-closed is not a gauge choice but a requirement of the constraints.
Now, equating the $[4,4]$ components on both sides at linear order, we get a dynamical equation of motion given by
\begin{align}
  &e^{4\Omega^{(0)}+4A^{(0)}}\hat{d}\hat{\star}\hat{d}u\wedge B_1\wedge\tilde{\star}^{(0)}G_3^{(0)} + e^{2\Omega^{(0)}}\hat{d}\hat{\star}\hat{d}u\wedge\tilde{\star}^{(0)}\eta_2 =0\,,
\end{align}
\emph{which we do not satisfy offshell.} Instead we identify the linear fluctuation in the EOM operator $E_8$ as
\begin{align}
  \delta_u E_8=e^{2\Omega^{(0)}}\hat{d}\hat{\star}\hat{d}u\wedge \Big\{ie^{2\Omega^{(0)}+4A^{(0)}} B_1\wedge G_3^{(0)} + \tilde{\star}^{(0)}\eta_2\Big\}\,.
\end{align}
Note that the terms proportional to $u$ canceled following the discussion of section \ref{s:staticmod}.

\paragraph{$\tau$ equation of motion:} Next, upon linearizing the $\tau$ equation of motion, we obtain the linear fluctuation in the EOM operator $\delta_u E_\tau$ as\footnote{As detailed in appendix \ref{app:FlucTermsEOMBian}.}
\begin{align}\label{eq:Etau}
  \delta_u E_\tau = \delta\tau e^{2\Omega^{(0)}-4A^{(0)}}\hat{d}\hat{\star}\hat{d}u\wedge\tilde{\epsilon}^{(0)}\, . %=0\,.
\end{align}

\subsection{Solving the constraints}
\label{sec:SolveConstr}

As detailed in section \ref{sec:GeoModISD}, we have a constraint 
\begin{align}
  \tilde{d}\tilde{\star}^{(0)}\tilde{d}\delta e^{-4A}-\tilde{d}\tilde{\star}^{(0)}V_1=\frac{i}{2\textrm{Im}\tau^{(0)}}\left(G_3^{(0)}\wedge \tilde{d}\bar{\eta}_2+\tilde{d}\eta_2\wedge\bar{G}_3^{(0)}\right)\,. \tag{\ref{eq:F5Bianchi06}}
\end{align}
%
%
% \begin{align}
%   \tilde{\nabla}^{\tilde{2}}\delta e^{-4A} -\tilde{\nabla}^{\tilde{m}}V_{m}=-\frac{\tilde{d}\eta_2\cdot \bar{G}_3^{(0)}+\tilde{d}\bar{\eta}_2\cdot G_3^{(0)}}{2\textrm{Im}\tau^{(0)}}\,. \label{eq:LapDelWarp}
% \end{align}
%
Operating with $\tilde{d}$ on \eqref{eq:F5Bianchi15} also leads to \eqref{eq:F5Bianchi06}. That is, the integrability condition for \eqref{eq:F5Bianchi15} is satisfied by the Poisson equation for the fluctuation of the warp factor, which we have solved in terms of Green's functions.

Separately, taking the Hodge dual of \eqref{eq:F5Bianchi15} w.r.t. the background metric yields
\begin{align}
  &-\frac{1}{2}e^{4A^{(0)}}\partial_{m}\delta e^{-4A} + \frac{1}{2} e^{4A^{(0)}}V_{m} + \frac{1}{2}e^{2\Omega^{(0)}+4A^{(0)}}\tilde{\nabla}^{\tilde{n}}(\tilde{d}B_1)_{mn} \nn\\
  &\qquad\qquad =\frac{ie^{4A^{(0)}}}{4\textrm{Im}\tau^{(0)}} \Big[\tilde{\star}^{(0)}(\eta_2\wedge\bar{G}_3^{(0)}-\bar{\eta}_2\wedge G_3^{(0)})\Big]_m\,. \label{eq:HodgeStarF5Bianchi15}
\end{align}
When we compare this to the $\mu m$-components of the Einstein equations \eqref{eq:mnEinFieldEqn}, we obtain another constraint
\begin{align}
  \tilde{\nabla}^{\tilde{n}}\delta\tilde{g}_{mn}=0\,, \label{eq:Constr.CovTransv.MetricFluc}
\end{align}
i.e., $\delta\t g_{mn}$ is covariantly transverse (with respect to the background covariant derivative).
It is important to note that this is \emph{not} a gauge choice but instead a requirement of gauge invariance.

Likewise, the $G_3$ equation of motion \eqref{eq:G3eom35} leads to the co-closedness of $\eta_2$ as a constraint:
\begin{align}
  \tilde{d}\tilde{\star}^{(0)}\eta_2=0\,. \label{eq:Eta2Coclosed}
\end{align}
Once again, what appears to be a possible gauge choice is a \emph{requirement} of the constraints.
We have already noted that therefore \eqref{eq:Eta2EqnDirect} becomes a Poisson equation for $\eta_2$. While it is sometimes illuminating to find $\eta_2$ using an alternative approach, there is in principle a solution for $\eta_2$ in terms of Green's functions.
%Nevertheless, we take an alternative approach to solve for $\eta_2$.

All that remains is to determine the compensator $B_m(y)$.
Satisfying the off-diagonal (i.e. $\mu\neq\nu$) components of \eqref{eq:munuEinFieldEqn} gives the constraint
\begin{align}
  \tilde{\nabla}^{\tilde{m}}B_m=e^{-2\Omega^{(0)}}\delta e^{-4A}\,. \label{eq:DivB1}
\end{align}
Now as a consequence of \eqref{eq:DivB1}, all the $(\mu,\nu)$ components of the linear order Einstein field equations \eqref{eq:munuEinFieldEqn} are automatically satisfied. Next, substituting \eqref{eq:Constr.CovTransv.MetricFluc} and \eqref{eq:DivB1} into the $\mu m$-components of the linear order Einstein equations \eqref{eq:mnEinFieldEqn}, we obtain\footnote{Using $\tilde{\nabla}^{\tilde{n}}(\tilde{d}B_1)_{mn}=-\tilde{\nabla}^{\tilde{2}}B_m+\tilde{\nabla}_m\tilde{\nabla}^{\tilde{n}}B_n$, due to the Ricci flatness of the background metric.}
\begin{align}
  &\tilde{\nabla}^{\tilde{2}}B_m =  e^{-2\Omega^{(0)}}V_{m}-\frac{ie^{-2\Omega^{(0)}}}{2\textrm{Im}\tau^{(0)}} \Big[\tilde{\star}^{(0)}(\eta_2\wedge\bar{G}_3^{(0)}-\bar{\eta}_2\wedge G_3^{(0)})\Big]_m\,. \label{eq:LapB1}
\end{align}
This equation also follows from the Bianchi identity \eqref{eq:F5Bianchi15}.

With this, the constraint on $B_1$ in \eqref{eq:LapB1} takes the form of a Poisson equation, given by%\footnote{As in terms of components, we have $i\Big[\tilde{\star}^{(0)}(\eta_2\wedge\bar{G}_3^{(0)}-\bar{\eta}_2\wedge G_3^{(0)})\Big]_m=\left(\eta_{2np}\bar{G}_3^{(0)}{}_m{}^{\widetilde{np}}+\bar{\eta}_{2np}G_3^{(0)}{}_m{}^{\widetilde{np}}\right)/2$.}
%
%\begin{align}
%  \tilde{\nabla}^{\tilde{2}}B_m=-\delta e^{-\Omega}\partial_m e^{-4A^{(0)}}+e^{-2\Omega^{(0)}}f_m\,.
%\end{align}
%
\begin{align}
  \tilde{\nabla}^{\tilde{2}}B_m=e^{-2\Omega^{(0)}}f_m\,,\label{eq:BPoisson}
\end{align}
where $f_m$ is precisely as defined in \eqref{eq:PoissonWDiv}.
The solution of the Poisson equation is 
\begin{align}
  B_n=e^{-2\Omega^{(0)}}\int d^6Y \sqrt{\tilde{g}_{(0)}(Y)}G^{\slashed{m}}{}_n(y,Y)f_{\slashed{m}}(Y)\label{eq:BGreen}
\end{align}
in terms of the bivector Green's function.
It is straightforward to verify using \eqref{eq:BiscaBivecRel} that the divergence $\tilde{\nabla}^{\tilde{n}}B_n$ is consistent with \eqref{eq:DivB1}.

Given a traceless, covariantly transverse geometric modulus $\delta\t g_{mn}$, we have now presented a formal solution for fluctuations in the warp factor (and therefore the 5-form), 3-form flux, axiodilaton, and the compensator $B_m(y)$ (in terms of Green's functions where appropriate).
The fact that the compensator \eqref{eq:BGreen} derived from the Poisson equation \eqref{eq:BPoisson} satisfies \eqref{eq:DivB1} is a nontrivial consistency check on our ansatz. 

% \paragraph{K\"ahler modulus:} \af{This all applies to static shifts, so it belongs in section 3.1.}
% For the case of a K\"ahler modulus as discussed in section \ref{sec:GeoModKahler}, the $(0,3)$ component of $W_3$ must vanish. Hence from the condition \eqref{eq:ISDG3} we get
% %
% \begin{align}
%   \delta\tau=0\,.
% \end{align}
% \af{This actually follows because $W_{(3,0}$ and $W_{1,2}$ vanish.}
% With this, $\chi_3$ becomes equal to $\tilde{d}\eta_2$, and this must be a (2,1) form given by\footnote{Since, the Hodge dual of a $(2,1)$ form w.r.t. a K\"ahler metric gives another $(2,1)$ form.}
% %
% \begin{align}
%   \chi_3=\partial\bar{\partial}\Lambda_1\,,
% \end{align}
% %
% where $\Lambda_1$ is a $(1,0)$ form. \rma{Do we check nonprimitivity of $\tilde{d}\eta_2+i\tilde{\star}^{(0)}\tilde{d}\eta_2$?} \af{It is AISD by definition, so only the nonprimitive part of $\del\bar\del\Lambda_{(1,0)}$ survives.}

% Now taking Hodge dual of \eqref{eq:PrimtivePertG3Lin} w.r.t. $\tilde{g}^{(0)}_{mn}$, and using that the background K\"ahler form $\tilde{J}_2^{(0)}$ is proportional to $\star^{(0)}(\tilde{J}_2^{(0)}\wedge \tilde{J}_2^{(0)})$ due to $\tilde{g}^{(0)}_{mn}$ being a K\"ahler metric \cite{Martucci:2016pzt, huybrechts2005complex}, we obtain
% %
% \begin{align}
%   \tilde{\nabla}^{\tilde{2}}\Lambda_1=\,.
% \end{align}
% %
% \rma{This entire paragraph above with necessary corrections has been moved to section 3.1.}

\section{Modulus action and K\"ahler potential}
\label{sec:ModuliActionMetric}

Here we derive the kinetic action for the K\"ahler moduli as well as flat directions in the complex structure/axiodilaton sector. 
We remind the reader that the 4D EFT at second order in moduli is given by
\begin{align}
  S=&\frac{1}{4\kappa^2}\bigg[\int d^{10}X\sqrt{-g}\delta g^{MN}\delta E_{MN}-\frac{1}{2\left(\textrm{Im}\tau^{(0)}\right)^2}\int\left(\delta\tau\delta\bar{E}_{\bar{\tau}}+\delta\bar{\tau}\delta E_\tau\right) \nn\\
  &\qquad\qquad -\int \delta C_4^\prime\wedge\delta E_6-\frac{1}{2\textrm{Im}\tau^{(0)}}\int \left(\delta A_2\wedge\delta\bar{E}_8+\delta\bar{A}_2\wedge\delta E_8\right)\bigg]\tag{\ref{eq:EFTaction}}
\end{align}
after integration over the internal dimensions. The action is off shell, so we do not set the dynamical EOM operators to zero; however, the constraints (some of which ensure the fluctuations remain on moduli space and some of which are required by gauge invariance) must vanish.

To connect to the K\"ahler potential for the K\"ahler moduli, we include the volume modulus $c(x)$ (the universally present K\"ahler modulus) and axions $b_I(x)$ that descend from $C_4$ (which complete the holomorphic coordinates of the moduli space in 4D SUGRA) following \cite{Frey:2008xw,Frey:2013bha} respectively. In the following, $\{\omega_2^I\}$ is a basis of harmonic (1,1) forms on the background CY normalized such that
\be \int \omega_2^I\wedge \tilde{\star}^{\0}\omega_2^J = 3\tilde{V}\delta^{IJ}\label{eq:norm2form}\ee
with $\omega_2^1 \equiv\t J_2^\0$.
As a result, $I,J,\cdots$ run over the K\"ahler moduli and axions; we use $A,B,\cdots$ to denote all unstabilized metric deformations (K\"ahler and complex structure moduli), while $a,b,\cdots$ will denote flat directions among the complex structure moduli only. 
We consider the moduli $u^A$ to be real variables as appropriate for K\"ahler moduli; we comment on complex variables for complex structure flat directions below.
We do not consider 4D scalars that descend from the 2-form potentials or D-brane degrees of freedom, but it is a straightforward generalization to include them using the results of \cite{Frey:2013bha,Cownden:2016hpf}. 
In the remainder of this section, we omit the subscript and superscript $\0$ for simplicity when referring to values of the background; any perturbations are given explicitly and have a $\delta$ symbol preceding (or have been defined previously as being part of the fluctuations, like $V_1$ or $\eta_2$). Our discussion will be similar to that in \cite{Cownden:2016hpf} as the calculation is similar.

It is useful to define several combined fluctuations, where $\bm{\delta}$ indicates the inclusion of 4D spacetime dependence summed over all the moduli. For the $y$-dependent components only, $\delta^A$ indicates a variation with respect to any metric deformation and $\delta^I$ is with respect to a K\"ahler metric or axion fluctuation. We define the total fluctuation in the warp factor and Weyl factor as
\be\label{eq:totalwarp} \bm{\delta} e^{-4A(x,y)} = c(x) +u_A(x) \delta^A e^{-4A(y)}\Rightarrow \bm{\delta} e^{-2\Omega} = c(x) .\ee
Like the metric moduli we have considered here, the volume modulus and $C_4$ axion fluctuations have a compensator of the form $\delta g_{\mu m} = \partial_\mu f(x) B_m(y)$. For shorthand, we take
\be\label{eq:generalB} \mathbf{B}_1(x,y)\equiv -c(x)\tilde{d} K(y) +u_A(x) B^A_1(y) +b_I(x) \check{B}^I_1(y),\ee
where $B^A_1$ is the compensator defined in section \ref{s:linearization} and $\check{B}^I_1$ is the compensator for the axion as given in \cite{Frey:2013bha}. This combined compensator satisfies
\be\label{eq:combinedB}
\t\Del^{\t m} \B_m = -c(x) e^{-4A} +e^{-2\Omega} \bm{\delta} e^{-4A} .\ee

\subsection{Moduli space metric}

We recall that the quadratic effective action consists of the fluctuation in each 10D field contracted with the linearized first variation of the action (ie, the unsimplified linearized equations of motion). We impose the constraints but must leave the dynamical equations of motion off shell since the action is an off-shell quantity. We consider the contribution from each 10D field separately. 
In the following, an integral over $d^4x$ or with a subscript 4 indicates integration over the Minkowski spacetime, an integral over $d^6y$ or with subscript 6 indicates integration over the CY manifold, and an integral over $d^{10}X$ or with no subscript is over the entire 10D spacetime.

\paragraph{$\tilde F_5$ contribution}

Including the axions, the fluctuation in the $C_4$ potential is
\be \bm{\delta} C'_4 = -e^{2\Omega} u_A(x)\tilde\star\tilde d B^A_1(y) +b_I(x)\tilde\star\omega_2^I(y) -\hat d b_I(x) \wedge K_3^I(y) ,\ee
where $K_3^I$ is a compensator for the axion. The dynamical EOM operator is
\be \bm{\delta} E_6 = e^{2\Omega} \hat d \hat\star\hat d b_I\wedge \gamma_2^I -e^{4\Omega} (\hat d \hat\star\hat d)\tilde d(e^{4A} \mathbf{B}_1) ,\ee
where $\gamma_2^I$ is a harmonic 2-form $\gamma_2^I=C^{IJ}\omega_2^J$ that solves the constraints for the axion. The matrix $C^{IJ}$, which depends on the position in moduli space, is defined in \cite{Frey:2013bha}.\footnote{We present the formula in \eqref{eq:metricKahler4} through $3\t V (C^{-1})^{IJ}=\mathcal{G}^{IJ}$.} 

After integration by parts (which eliminates some terms), the contribution to the effective action from the 5-form sector is
\be\label{eq:F5action1}
S_5 &= -\frac{1}{4\kappa^2}\int \bm{\delta} C'_4\wedge \bm{\delta} E_6 = -\int_4 e^{2\Omega} b_I \wedge \hat d\hat\star\hat d b_J \int_6 \tilde\star \omega_2^I\wedge\gamma_2^J-\int e^{6\Omega} u_A \tilde d\tilde\star\tilde d B_1^A\wedge(\hat d\hat\star\hat d)(e^{4A} \mathbf{B}_1) .
\ee
Using equations \eqref{eq:norm2form} and \eqref{eq:F5Bianchi15} and converting to index notation, we find
\be\label{eq:F5action2}
S_5 =& \frac{3\tilde V}{4\kappa^2} C^{IJ} \int d^4x e^{2\Omega} b_I\hat\del^{\hat 2}b_J
+\frac{1}{4\kappa^2}\int d^4x\int d^6y\sqrt{\tilde g}  e^{4\Omega}e^{4A} u_A\hat\del^{\hat 2} \B^{\t m}\left[\del_{m}(\delta^Ae^{-4A})
-V^A_m\vphantom{\frac 12}\right.\\
&\left. +\frac{i}{2\textrm{Im}\tau} \t\star\left(\eta_2^A\wedge \bar G_3-\bar\eta_2^A\wedge G_3\right)_m \right] .
\ee

\paragraph{Einstein contribution}
The only dynamical EOM among the Einstein equations is the $(m,n)$ component, 
\be\label{eq:einsteinEOM}
\bm{\delta} E_{mn} =& \hat\del^{\hat 2} \left[\frac 12 \bm{\delta} e^{-4A} e^{-2\Omega}\t g_{mn}  -\frac 32 c e^{-4A}\t g_{mn}-\frac 12 u_A e^{-2\Omega}
e^{-4A}\delta^A\t g_{mn}+\t\Del_{(m}\B_{n)} +4\del_{(m}\B_{n)}\right.\\
&\left. -\t\Del^{\t p}\B_p\, \t g_{mn} -2\del^{\t p}A\B_p\, \t g_{mn}\right] ,
\ee
so we are only concerned with the fluctuation
\be\label{eq:invmetricfluct}
\bm{\delta} g^{mn} = -\frac 12 e^{6A}\t g^{mn}\bm{\delta} e^{-4A} -e^{2A} u_A \t g^{mp}\t g^{nq} \delta^A\t g_{pq} .
\ee
The contribution of the gravitational sector to the quadratic action simplifies considerably with equation \eqref{eq:combinedB}, yielding
\be\label{eq:EHaction1}
S_g =&\frac{1}{4\kappa^2} \int d^{10}X\sqrt{|g|}\bm{\delta} g^{mn}\bm{\delta} E_{mn} =\frac{1}{4\kappa^2} \int d^4x\int d^6y\sqrt{\t g} e^{4\Omega} \left\{ 3\bm{\delta} e^{-4A} \hat\del^{\hat 2} c -e^{4A}\del^{\t p}(\bm{\delta} e^{-4A})
\hat\del^{\hat 2} \B_p\vphantom{\frac 12}\right.\\
&\left. +\frac 12 e^{-2\Omega}e^{-4A}(u_A\hat\del^{\hat 2}u_B) \t g^{mp}\t g^{nq}\delta^A\t g_{mn}\delta^B\t g_{pq}
-u_A\delta^A\t g_{mn} \hat\del^{\hat 2} \left[\t\Del^{\t m}\B^{\t n} +4\del^{\t m}A\B^{\t n}\right]
\right\} .
\ee
We can write $4\delta^A\t g_{mn}\del^{\t m}A=-e^{4A}V_n^A$ in the last term, and the penultimate term vanishes on integration by parts over $y$ due to the constraint $\t\Del^{\t m}\delta^A\t g_{mn}=0$. Furthermore, the $y$ integral of $\bm{\delta} e^{-4A}$ in the first term gives $\t V c$ because we have seen that $\delta^A e^{-4A}$ integrates to zero.

\paragraph{$G_3$ contribution}
The fluctuation of the 2-form potential is $\bm{\delta} A_2 = u_A\eta_2^A-\hat d b_I\wedge\Lambda_1^I$, where $\Lambda_1^I$ is a compensator needed to describe the axionic degrees of freedom (and is in fact equal to $\Lambda_1$ as defined in section \ref{sec:GeoModKahler}).
The associated EOM operator is
\be\label{eq:E8operator}
\bm{\delta} E_8&=e^{2\Omega}\hat d\hat\star\hat d u_A\wedge\t\star\eta_2^A+e^{2\Omega}\hat d\hat\star\hat d b_I\wedge\t\star\t d\Lambda_1^I +ie^{4\Omega}e^{4A}\hat d\hat\star\hat d\B_1\wedge G_3 .
\ee
The contribution to the action is
\be\label{eq:G3action1}
S_3 =& -\frac{1}{8\kappa^2\mathrm{Im}\,\tau} \int e^{2\Omega} \left\{ u_A\eta_2^A\wedge \left[ \hat d\hat\star\hat d u_B\wedge\t\star\bar\eta_2^B +
\hat d\hat\star\hat d b_J\wedge\t\star\t d\bar\Lambda_1^J-ie^{2\Omega}e^{4A}\hat d\hat\star\hat d \B_1\wedge\bar G_3\right]+\mathrm{c.c.}\right\} .
\ee
The second term vanishes after integration by parts due to the constraint $\t d\t\star\eta_2^A=0$. With some simplification,
\be\label{eq:G3action2}
S_3 =& \frac{1}{8\kappa^2\mathrm{Im}\,\tau} \int d^4x e^{2\Omega} u_A\hat\del^{\hat 2} u_B \int_6 \left(\eta_2^A\wedge\t\star\bar\eta_2^B+\bar\eta_2^A\wedge\t\star\eta_2^B\right)\\
& -\frac{i}{8\kappa^2\mathrm{Im}(\tau)} \int d^4x\int d^6y\sqrt{\t g} e^{4\Omega}e^{4A} u_A\hat\del^{\hat 2} \B^{\t m} \t\star\left(\eta_2^A\wedge \bar G_3-\bar\eta_2^A\wedge G_3\right)_m .
\ee

\paragraph{Axiodilaton contribution}
As we have seen, the axiodilaton is $\tau(x,y) = \tau^\0 + u_A(x)\delta^A\tau$, and the EOM operator is given by \eqref{eq:Etau}. Again dropping the $\0$ superscript for convenience, the
contribution to the action is
\be\label{eq:Stau}
S_\tau = & -\frac{1}{8\kappa^2(\mathrm{Im}\,\tau)^2} \int \left(\bm{\delta}\tau \bm{\delta}\bar E_\tau+\bm{\delta}\bar\tau\bm{\delta} E_\tau\right) \\
=& -\frac{1}{8\kappa^2(\mathrm{Im}\,\tau)^2} \int_4 e^{2\Omega} u_A\hat d\hat\star\hat d u_B\left(\delta^A\tau\delta^B\bar\tau +\delta^A\bar\tau \delta^B\tau\right)\int_6 e^{-4A}\t\epsilon\\
=& \frac{1}{8\kappa^2(\mathrm{Im}\,\tau)^2} \t V \int d^4x\, u_A\hat\del^{\hat 2} u_B\left(\delta^A\tau\delta^B\bar\tau +\delta^A\bar\tau \delta^B\tau\right)
\ee
following \eqref{eq:defOmega}.

\paragraph{Kinetic action}

The kinetic action for the moduli we consider is therefore $S=S_5+S_g+S_3+S_\tau$. As it turns out, all the terms containing $\B_1$ cancel in the sum (once we recognize that 
$\del_m (\bm{\delta} e^{-4A}) = u_A\del_m(\delta^A e^{-4A})$), so we are left with
\be\label{eq:kineticaction}
S =& \frac{3\t V}{4\kappa^2} \int d^4x\, e^{2\Omega}\left( e^{2\Omega} c\hat\del^{\hat 2} c + C^{IJ} b_I\hat\del^{\hat 2} b_J +\frac{1}{3\t V} \mathcal{G}^{AB}u_A\hat\del^{\hat 2}u_B\right) ,
\ee
where the metric $\mathcal{G}^{AB}$ is given by 
\be\label{eq:metricmodulispace} \mathcal{G}^{AB} = &
\frac 12 \int d^6y\sqrt{\t g}\, \left(e^{-4A} \t g^{mp}\t g^{nq}\delta^A\t g_{mn}\delta^B\t g_{pq} \right) + \frac{1}{2\mathrm{Im}\,\tau} \int_6 \left(\eta_2^A\wedge\t\star\bar\eta_2^B+\bar\eta_2^A\wedge\t\star\eta_2^B\right)\\
& +\frac{\t V}{2(\mathrm{Im}\,\tau)^2} e^{-2\Omega} \left(\delta^A\tau\delta^B\bar\tau +\delta^A\bar\tau \delta^B\tau\right) .
\ee
We discuss features of this metric for different moduli below.
In \eqref{eq:metricmodulispace}, $\Omega$, $C^{IJ}$, and $\mathcal{G}^{AB}$ are evaluated at the fixed GKP background, so they are strictly speaking constants, and the action is purely quadratic in the fluctuations of the moduli. We can therefore freely integrate by parts to write it as a sum over squared first derivatives.
However, our analysis is valid for any GKP background, so we can promote $\Omega$, $C^{IJ}$, and $\mathcal{G}^{AB}$ to functions of the moduli space. These functions yield the metric on moduli space.

In the limit of very large volume, the flux energy density dilutes and the warp factor becomes trivial.
In that limit, $e^{-4A}$ approaches the constant value $e^{-2\Omega}$, which becomes large. In \eqref{eq:metricmodulispace}, the $\eta_2$ terms, which are defined with respect to the background metric, remain fixed in size while the other two terms grow, and $\delta\t g_{mn}$ terms become the same integrals as in an unwarped product compactification. 
Due to the overall factor of $e^{2\Omega}$ multiplying $\mathcal{G}^{AB}$, the kinetic action then goes to the usual kinetic action for the moduli space of a CY compactification with corrections due to the small $y$ dependence of the warp factor and the $\eta_2$ terms.
The flux quantization conditions imply that both of these corrections are of order $(\alpha')^2$ corresponding to their suppression by the (dimensionless) inverse warped volume $e^{2\Omega}$.
On the other hand, the $\delta^A\tau$ terms continue to contribute at large volume because they indicate the orientation of the complex structure flat directions in the field space of the 10D supergravity.
Finally, it is important to remember that strongly warped throat regions can exist even in compactifications that are large enough for curvature corrections to be small everywhere; in that case, the warp factor and flux are not simply corrections to the moduli space metric of a CY compactification.

\subsection{K\"ahler metric for K\"ahler moduli}
To consider K\"ahler moduli (other than the volume modulus), we restrict to metric deformations $u_I\delta^I\t g_{mn}$ described by the harmonic (1,1) forms $\omega_2^I$ with $I\geq 2$.
We can therefore connect to the language of differential forms as typically used to describe the moduli space of CY manifolds \cite{Candelas:1990pi}.

\paragraph{Metric on K\"ahler moduli space:}
There are several simplifications in \eqref{eq:metricmodulispace} in this case. The first to note is that $\delta^I\tau=0$. As we have noted, K\"ahler moduli deformations in GKP compactifications do not shift the axiodilaton.

In terms of complex coordinates, the first integral in \eqref{eq:metricmodulispace} is
\be\label{eq:metricKahler1}
\frac 12 \int d^6y\sqrt{\t g}\, \left(e^{-4A} \t g^{mp}\t g^{nq}\delta^I\t g_{mn}\delta^J\t g_{pq} \right)
=&\int d^6y\sqrt{\t g}\, \left(e^{-4A} \t g^{i\bar\jmath} \t g^{\bar a b} (-i\omega_{i\bar a})^I(i\omega_{\bar\jmath b}^J)\right)\\
=&\int_6 e^{-4A}\omega_2^I\wedge\t\star\omega_2^J\, .
\ee
The factor of 2 in the first line is because each real index runs over both holomorphic and antiholomorphic values.

The final terms in \eqref{eq:metricmodulispace} are given by 
\be\label{eq:metricKahler2}
\int_6 \eta_2^I\wedge\t\star\bar\eta_2^J
= \int_6\left( \bar\del\Lambda_1^I-\del\Lambda_1^J\right)\wedge
\t\star\left(\del\bar\Lambda_1^I-\bar\del\bar\Lambda_1^J\right)
=\int_6\left( \bar\del\Lambda_1^I\wedge\t\star\del\bar\Lambda_1^J
+\del\Lambda_1^I\wedge\t\star\bar\del\bar\Lambda_1^J\right)
\ee
based on counting holomorphic and antiholomorphic indices.
After integrating by parts, we find
\be\label{eq:metricKahler3}
\int_6 \eta_2^I\wedge\t\star\bar\eta_2^J
= \int_6\left( \Lambda_1^I\wedge\bar\del\t\star\del\bar\Lambda^J_1+\Lambda_1^I\wedge\del\t\star\bar\del\bar\Lambda^J_1\right)
=i\int_6 \Lambda_1^I\wedge \bar G_3\wedge\omega_2^J\, .
\ee
All told, we find that
\be\label{eq:metricKahler4}
\mathcal{G}^{IJ} =\int_6 e^{-4A}\omega_2^I\wedge\t\star\omega_2^J
+\frac{i}{2\mathrm{Im}\,\tau}\int_6\left(\Lambda_1^I\wedge \bar G_3-\bar\Lambda_1^I\wedge G_3\right)\wedge\omega_2^J
\ee
for $I,J\geq 2$. 
This is precisely the matrix $3\t V (C^{-1})^{IJ}$ defined in \cite{Frey:2013bha}; note that our $\Lambda_1^I$ satisfies the same Poisson equation as $\check{\Lambda}_1^I$ as defined in \cite{Frey:2013bha}, so it is the same 1-form.
While we have so far defined $\mathcal{G}^{IJ}\equiv 3\t V (C^{-1})^{IJ}$ only for $I,J\geq 2$, it is also sensible to evaluate it for $I=1$ by recalling that $\omega_2^1=\t J_2$, the K\"ahler form of the background, in which case $(C^{-1})^{1J}$ simplifies considerably \cite{Frey:2013bha}.
First, we recall that $G_3\wedge\t J_2=0$, also implying $\Lambda_1^1=0$, so the second integral in \eqref{eq:metricKahler4} vanishes for either $I=1$ or $J=1$.
Further, constancy of the pointwise inner product of $\t J_2$ with any 2-form implies that the first integral is $(\t J_2\cdot\omega_2^I)e^{-2\Omega}\t V=3e^{-2\Omega}\t V\delta^{1I}$ with the normalization \eqref{eq:norm2form}.

We note that $\mathcal{G}^{IJ}=3\t V\delta^{IJ}$ for any compactification with formal metric, including those with torus factors.
In that case, $\omega_2^I\wedge\t\star\omega_2^J=3\t V\delta^{IJ} \t\epsilon$, so the geometric term in \eqref{eq:metricKahler4} gives $3\t V\delta^{IJ}e^{-2\Omega}$, and the flux compensators vanish because $G_3\wedge\omega_2^I$ is a harmonic 5-form. These do not exist on a CY, and we have already assumed that this product vanishes on a torus compactification by restricting to massless moduli.

\paragraph{K\"ahler potential and no-scale structure:}
In a 4D EFT, the moduli can be organized into complex scalars and the metric on moduli space is the second derivative of a K\"ahler potential.
Following \cite{Martucci:2016pzt}, the holomorphic coordinates of the 4D supergravity are $\rho_1 \equiv c+ib_1$ along with 
\be\label{eq:holomorphicSUGRA}
\rho_I = (C^{-1})^{IJ}u_J+i b_I
\ee
for $I,J\geq 2$ for linear fluctuations around a fixed point in moduli space (i.e., fixed background metric, warp factor, etc).
Therefore, the action on the K\"ahler moduli space altogether becomes
\be\label{actionKahler}
S = -\frac{3\t V}{4\kappa^2} \int d^4x\, e^{2\Omega} C^{IJ} \del_\mu\rho_I\del^\mu\bar\rho_J\, ,
\ee
where the sum runs over $I,J=1,\ldots h_{1,1}$ because $C^{1I}=e^{2\Omega}\delta^{1I}$ (as we saw above).
The final action \eqref{actionKahler} agrees with \cite{Martucci:2016pzt}.

To promote $C^{IJ}$ to the metric on moduli space and write it in terms of a K\"ahler potential, we can extend the definition of the $\rho_I$ to finite deformations following \cite{Martucci:2016pzt}.
Then the action \eqref{actionKahler} follows from a K\"ahler potential $K=-3\ln(e^{-2\Omega})$ with the dimensionless warped volume $e^{-2\Omega}$ understood as a function of all the K\"ahler moduli \cite{Martucci:2016pzt}. 
It is not possible to write $K$ explicitly in terms of the holomorphic variables $\rho_I$ because the real part of \eqref{eq:holomorphicSUGRA} is not explicitly invertible due to the dependence of $C^{IJ}$ on the background K\"ahler moduli; however, it is possible to differentiate $K$ implicitly.
When evaluated on the initial background in our basis, $\del_I K=0$ except for $I=1$ corresponding to the volume modulus (as we have seen $\delta e^{-2\Omega}=0$ except for the volume modulus), but the second derivatives need not vanish because the direction of increasing volume changes with position in moduli space.

Because GKP solutions at the 10D level always have a massless volume modulus and Minkowski spacetime, the 4D SUGRA has a no-scale structure. 
The no-scale structure is a nontrivial condition on the K\"ahler potential for the K\"ahler moduli, which must hold even including corrections from the flux and warp factor. 
Using implicit derivatives of $K$, \cite{Martucci:2016pzt} showed that $K$ is in fact no-scale.

\subsection{Complex structure flat directions}\label{s:kineticcomplex}

The metric $\mathcal{G}^{AB}$ of \eqref{eq:metricmodulispace} applies to complex structure flat directions as well as the K\"ahler moduli. 

\paragraph{Possible mixing with K\"ahler moduli:}
Before considering the metric $\mathcal{G}^{ab}$, we also consider whether there is mixing $\mathcal{G}^{Ia}$ between the two sectors, K\"ahler and complex structure.
Because the K\"ahler moduli do not have associated axiodilaton shifts $\delta^I\tau$, the final term in \eqref{eq:metricmodulispace} vanishes. Furthermore, because the metric deformations $\delta^I\t g_{mn}$ and $\delta^a\t g_{mn}$ have different structure in terms of holomorphic and antiholomorphic indices, the integral over $\delta^I\t g_{mn}\delta^a\t g^{\t m\t n}$ also vanishes. 
However, it is not clear whether the integral over $\eta_2^I\wedge\t\star\bar\eta_2^a$ (or its conjugate) vanishes. We have seen that $\eta_2^I$ has $(1,1)$ and $(2,0)$ components. 
Similarly, we have seen that $\eta_2^a$ may have $(2,0)$, $(1,1)$, and $(0,2)$ components (taking both $\eta_2^{\pm a}$ for a real flat direction). 
As a result, it is not clear whether the required shift in the 3-form flux induces a mixing between K\"ahler and complex structure moduli. This would be a qualitatively new feature due entirely to the presence of background $G_3$ flux. 
As we have noted previously, these terms are suppressed at large volume, so any mixing vanishes at infinite volume.

\paragraph{Complex structure flat direction metric:}
We now turn to the metric for the complex structure flat directions themselves. 
For a single real flat direction, $\mathcal{G}^{ab}$ takes the general form \eqref{eq:metricmodulispace} with $\delta\t g_{mn}$ given by a linear combination of $\delta^\pm\t g_{mn}$. 
In the remainder of this section, we consider complex flat directions in terms of complex variables.
These results follow either from writing the action for real and imaginary parts of the moduli and combining them into complex fields or from a careful calculation directly in terms of complex variables.

We can simplify the first term of \eqref{eq:metricmodulispace} involving the metric deformations. In fact, using \eqref{eq:OmegaIdent}, the contraction of the metric deformations is the same as in the result of \cite{Candelas:1990pi} in terms of the wedge product of harmonic $(2,1)$ forms and their conjugates. 
For complex-valued moduli fields, we find\footnote{Note that harmonic $(2,1)$ and $(1,2)$ forms on CY are ISD and AISD respectively.}
\be\label{eq:complexaction}
\frac 12 \int d^6y\sqrt{\t g}\, \left(e^{-4A} \t g^{mp}\t g^{nq}\delta^{+a}\t g_{mn}\delta^{-\bar b}\t g_{pq} \right) = 4i\int_6 e^{-4A} \xi^{a}_3\wedge \bar\xi^{\bar b}_3
\ee
with the appropriate sum of terms for the case of real-valued moduli fields. This is the classical result for CY compactifications \cite{Candelas:1990pi} modified by the warp factor. 
The flux terms give integrals of $\eta^{\pm a}_2\wedge\t\star\bar\eta^{\pm \bar b}_2$; however, possible integrals of $\eta^{\pm a}_2\wedge\t\star\bar\eta^{\mp \bar b}_2$, which could generate kinetic terms for $\del_\mu u^a\del^{\hat\mu}u^b$ and $\del_\mu \bar u^{\bar a}\del^{\hat\mu}\bar u^{\bar b}$, vanish by counting holomorphic and antiholomorphic indices since $\eta^+_2$ is $(1,1)$ and $(0,2)$ while $\eta^-_2$ is $(2,0)$. In all, the action and moduli space metric are
\begin{align}
S =& -\frac{2}{4\kappa^2} \int d^4x\, e^{2\Omega} \mathcal{G}^{a\bar b} \del_\mu u^a \del^{\hat\mu}\bar u^{\bar b}\,,\label{eq:actioncpx} \\
\mathcal{G}^{a\bar b} =& 2i\int_6 e^{-4A} \xi_3^a\wedge\bar\xi_3^{\bar b} +\frac{1}{2\mathrm{Im}\,\tau} \int_6 \left( \eta^{+ a}_2\wedge\t\star\bar\eta_2^{+\bar b} + \bar\eta_2^{-a}\wedge\t\star \eta_2^{-\bar b}\right) +\frac{\t V}{2(\mathrm{Im}\,\tau)^2} \delta^+\tau^a\delta^+\bar\tau^{\bar b} .\label{eq:metriccpx}
\end{align}
The factor of 2 in \eqref{eq:actioncpx} accounts for the term with $\mathcal{G}^{\bar ba}$.
The flux and axiodilaton terms appear to be idiosyncratic and particular to the given flat directions, so we do not have a simplified form. It is worth noting that both of these terms can mix different flat directions.
It would be very interesting to work out $\mathcal{G}^{a\bar b}$ in detail for explicit examples, such as those given by \cite{Cicoli:2022vny}.

Note that we have assumed that $\delta_u\tau=u\delta^+\tau$ with no $\bar u\delta^-\tau$ contribution. This follows from \eqref{eq:ISDG3full} for unbroken SUSY ($\lambda=0$), when we typically expect to find complex rather than real flat directions. When SUSY is broken, the flat direction can be complex only if $W_3^+$ \eqref{eq:CompW2} has no harmonic part, in which case $\delta^+\tau=0$ and $\delta^-\tau\neq 0$. In that case,
replace $\delta^+\tau\to\delta^-\tau$ in \eqref{eq:metriccpx}.

\paragraph{Condition on K\"ahler potential from ISD condition:}
An important feature of the 10D GKP solution is that the 3-form satisfies the ISD condition $\t\star G_3 = iG_3$ along the entire moduli space.
When translated into the language of 4D SUGRA, the ISD condition is given by the F-flatness conditions $D_a W=D_\tau W=0$, where $a$ is the index of a basis of complex structure moduli, $D=\del +\del K$ is the K\"ahler covariant derivative for K\"ahler potential $K$, and $W$ is the superpotential of Gukov-Vafa-Witten form \cite{Gukov:1999ya}
\begin{equation}
W = \int_6 G_3\wedge\tilde\Omega_3 .
\end{equation}
From the 4D perspective, solving the F-flatness conditions for the complex structure and axiodilaton gives our background GKP solution; if these equations have a continuous family of solutions, there are complex structure flat directions (alternatively, we can also find flat directions by perturbing around a single fixed solution, which is our approach in section \ref{sec:GeoModComplex}).

Under variation of complex structure moduli, $\del_a \t\Omega_3 =k_a\t\Omega_3 +\xi^a_3$, where $k_a$ is a function of the complex structure moduli space (i.e., the background metric) but not the coordinates and $\xi^a_3$ is the harmonic $(2,1)$ form associated with the complex structure modulus.
Meanwhile, $\del_a G_3$ is exact (and is often treated as vanishing, as e.g. in \cite{Giddings:2001yu}). Therefore, the F-flatness conditions are
\begin{align}
D_a W =& \int_6 G_3\wedge \xi^a_3 +\left(\del_a K +k_a\right)W=0\,, \nonumber\\
D_\tau W =& -\int_6 \frac{\bar G_3 \wedge\t\Omega_3}{\tau-\bar\tau} +\left( \del_\tau K +\frac{1}{\tau-\bar\tau}\right) W =0\,. \label{eq:fflat}
\end{align}
The first terms on the right-hand-side of each line of \eqref{eq:fflat} respectively vanish when $G_3$ has no $(1,2)$ and no $(3,0)$ components, which implies that the flux is ISD (assuming primitivity\footnote{This automatically holds for CY, as harmonic forms are also primitive.}). Therefore, the F-flatness conditions reproduce the ISD condition on $G_3$ if the last terms vanish. 
That occurs automatically if the superpotential $W=0$ on the background, which is also the F-flatness condition for the K\"ahler moduli. 
Otherwise, requiring \eqref{eq:fflat} to match the ISD condition is a condition on the K\"ahler potential.

For a product CY compactification, the K\"ahler potential is \cite{Candelas:1990pi}
\begin{equation}
K_{CY}(u^a,\bar u^a,\tau,\bar\tau) = -\ln [-i(\tau-\bar\tau)] - \ln\left[ -i\int_6 \t\Omega_3\wedge\bar{\t\Omega}_3\right] ,
\end{equation}
which is the form assumed by \cite{Giddings:2001yu}. With this form, $\del_\tau K_{CY} = -1/(\tau-\bar\tau)$ and 
\begin{equation}
\del_a K_{CY} = -\frac{\int_6\del_a\t\Omega_3\wedge\bar{\t\Omega}_3}{\int_6 \t\Omega_3\wedge\bar{\t\Omega}_3} = -k_a
\end{equation}
because $\xi_3^a\wedge\bar{\t\Omega}_3=0$. As a result, the F-flatness conditions always lead to the ISD condition.

The same is true for a K\"ahler potential of DeWolfe-Giddings form \cite{DeWolfe:2002nn}
\begin{equation}
K_{DWG}(u^a,\bar u^a,\tau,\bar\tau) = -\ln [-i(\tau-\bar\tau)] - \ln\left[ -i\int_6 e^{-4A}\t\Omega_3\wedge\bar{\t\Omega}_3\right] .
\end{equation}
As noted by \cite{Douglas:2007tu}, $\del_a K_{DWG}=-k_a$ also because the integral of $\del_a e^{-4A}$ times the volume form $-i\t\Omega_3\wedge\bar{\t\Omega}_3$ vanishes. While \cite{Douglas:2007tu} assumed this latter fact, we have demonstrated it explicitly.
If we further assume that the integral of the second variation $\del_a\del_{\bar b}e^{-4A}$ vanishes, the second derivative of $K_{DWG}$ also yields all but the flux compensator terms in \eqref{eq:metriccpx} (see \cite{Douglas:2007tu}); whether the DeWolfe-Giddings K\"ahler potential can accommodate these terms (possibly with a change of holomorphic variables in the 4D theory) is an important question. 
If not, the question is then how a further modified K\"ahler potential can both generate the kinetic terms we have found and the ISD conditions from F-flatness.

Of course, our results directly address the K\"ahler potential for flat directions only, rather than for the complex structure moduli (and axiodilaton) as a whole.
However, the standard assumption is that the K\"ahler metric and potential of the flat directions are just restrictions of those quantities from the full geometric moduli space to the flat directions, so our results are related.
As we have seen, the ISD condition automatically follows from F-flatness regardless of the K\"ahler potential if the superpotential vanishes; our work also indicates that $\del_a K=-k_a$ for a torus compactification since $\eta_2=0$ in that case, so $K=K_{DWG}$ holds (this seems likely true for any compactification with a formal metric).
For the 10D and 4D theories to match, the ISD condition \eqref{eq:ISDG3full} should follow from linearization of $D_a W=D_\tau W=0$ around a fixed background (these are second variations of the superpotential). 
Let us consider these conditions for fluctuations $(\delta\tau,\delta\bar\tau,\xi_3,\bar\xi_3)$ assuming that $\del_a K = -k_a$ and $\del_\tau K =-1/(\tau-\bar\tau)$ identically.

Start by considering
\begin{align}
\delta(D_\tau W) =& \int \frac{\delta\tau-\delta\bar\tau}{(\tau-\bar\tau)^2} \bar G_3\wedge\tilde\Omega_3 -\int \frac{\delta\bar\tau}{(\tau-\bar\tau)^2} (G_3-\bar G_3)\wedge\t\Omega_3
-\frac{1}{(\tau-\bar\tau)}\int \bar G_3\wedge \delta\t\Omega_3\nonumber\\
=& \frac{\delta\bar\tau}{4(\mathrm{Im}\,\tau)^2}\int G_3\wedge\t\Omega_3 +\frac{i}{2\mathrm{Im}\,\tau}\int \bar G_3\wedge\xi_3
\end{align}
since $D_\tau W=0$. With $G_3 = \xi_G+\lambda\t\Omega_3$,
\begin{equation}
\delta(D_\tau W) = \frac{i}{4\mathrm{Im}\,\tau} \int\left( 2\bar\xi_G\wedge\xi_3 +\frac{i\lambda\delta\bar\tau}{\mathrm{Im}\,\tau}\t\Omega_3\wedge\bar{\t\Omega}_3\right) 
= \frac{i}{4\mathrm{Im}\,\tau} \int \left( 2\xi_3\cdot\bar\xi_G +\frac{i\lambda\delta\bar\tau}{\mathrm{Im}\,\tau}\right)\t\Omega_3\wedge\bar{\t\Omega}_3 .
\end{equation}
For a flat direction, this must vanish, which is the same as the conjugate of \eqref{eq:ISDG3full} integrated against $\t\Omega_3$ (see \eqref{eq:CompW1}).

Similarly, after simplification,
\begin{equation}
\delta(D_a W ) = \int\left( G_3\wedge\delta\xi_3^a -\frac{i\delta\tau}{2\mathrm{Im}\,\tau}\bar\xi_G\wedge\xi_3^a  \right) ,\label{eq:deltaDaW}
\end{equation}
where $\delta\xi_3^a$ is the variation of the basis $(2,1)$ form with a fluctuation of complex structure associated with $\xi_3$.
Because $G_3$ has only $(2,1)$ and $(0,3)$ components, we need to know only the $(1,2)$ and $(3,0)$ components of $\delta\xi_3^a$.
Following the argument of \cite{Candelas:1990pi} for the variation of $\t\Omega_3$, we find 
\begin{equation}
(\delta\xi^a)_{ijk} = -\frac{3}{2} (\xi^a)_{[ij}{}^b \delta^-\t g_{k]b}\quad ,\quad (\delta \xi^a)_{i\bar\jmath\bar k} = -\frac{1}{2} (\xi^a)_{i[\bar\jmath}{}^{\bar b}\delta^+\t g_{\bar k]\bar b}\, ,
\end{equation}
which is strikingly similar to the definition of $W_3$. 
In fact, with this variation, it is straightforward to see that $\delta(D_a W)=0$ as given by \eqref{eq:deltaDaW} is equivalent to the integral of the wedge product of \eqref{eq:ISDG3full} with $\xi_3^a$ (using (\ref{eq:CompW2},\ref{eq:CompW1}) for the $(1,2)$ components of $W_3$).

As a result, insisting that F-flatness of the 4D theory reproduce the ISD condition for a complex structure flat direction places a consistency condition on the K\"ahler potential for the flat direction and the relationship between 10D moduli and the holomorphic variables of the 4D SUGRA. 
It is important to note that the superpotential itself should not be perturbatively renormalized \cite{Burgess:2005jx}.
There is an alternative scenario. As we discussed above, GKP compactifications have a no-scale structure, which is typically written as a constraint on the K\"ahler potential for the K\"ahler moduli.
However, we have seen that there are possibly cross terms in the moduli space metric between the K\"ahler moduli and complex structure flat directions. 
Therefore, it is possible that these terms violate the no-scale structure in such a way that is precisely compensated by nonvanishing of $\del_a K +k_a$ and $\del_\tau K +1/(\tau-\bar\tau)$ along the flat directions.
This resolution seemingly requires a coincidence between the values of $\eta_2^I$ and $\eta^{\pm a}_2$ for the K\"ahler moduli and flat directions.

\section{Discussions}
\label{sec:Discuss}

In this paper, we have given a first-principles derivation of the 4D kinetic action and moduli space metric for the geometric K\"ahler and flat complex structure moduli of GKP-type flux compactifications by dimensionally reducing the 10D type IIB supergravity action. 
To ensure that the modes we consider remain on the moduli space of the GKP compactification, we impose constraints on the fluctuations of the 10D metric and form fields such that $G_3$ remains ISD with respect to the perturbed metric. An important feature is that the warp factor receives correction from CY metric deformations, including from the perturbation of $G_3$. 
The fluctuation of 3-form flux $G_3$ is given in terms of a 2-form potential $\eta_2$, which we find in terms of the CY metric fluctuations and the background 3-form flux. This verifies that all geometric K\"ahler moduli are moduli of the full compactification; some complex structure deformations can also remain flat, possibly in combination with the axiodilaton.
This behavior has been reported earlier, e.g. in \cite{Demirtas:2019sip, Broeckel:2021uty, Cicoli:2022vny}, based on arguments involving the 4D superpotential; we see it here directly from the 10D equations of motion.

When the moduli are allowed to vary over the 4D spacetime, we introduce a linear compensator in the 10D metric and additional linear components in the form fields, as captured in our ansatzae \eqref{eq:metricansatz} and \eqref{eq:formsansatz}. We detail the constraints placed on these fluctuations to satisfy the 10D linearized supergravity equations of motion and Bianchi identities. 
%The complete set of constraints on the $y$-dependence of the 10D fluctuations is given by \eqref{eq:LapDelWarp}, \eqref{eq:Constr.CovTransv.MetricFluc}, \eqref{eq:DivB1}, \eqref{eq:LapB1}, and \eqref{eq:Eta2Coclosed}. 
We provide explicit solutions for \eqref{eq:LapDelWarp} and \eqref{eq:LapB1}, which describe the warp factor fluctuation $\delta e^{-4A}$ and the metric compensator $B_m$, respectively, in terms of biscalar and bivector Green's functions. We also use their properties to demonstrate consistency with \eqref{eq:DivB1}. 
Two other constraints are important. One is that the potential $\eta_2$ is co-closed, which helps determine its value.
Notably, the CY metric fluctuation is constrained to be covariantly transverse \eqref{eq:Constr.CovTransv.MetricFluc} with respect to the background CY metric. This relationship has sometimes been assumed earlier as a gauge choice for static deformations in the literature, but it is a constraint rather than a choice.

By imposing the aforementioned constraints off-shell, we derive the 4D moduli space metric for the geometric K\"ahler moduli and complex structure flat directions in \eqref{eq:metricmodulispace}. 
Unlike in the case of unwarped CY compactifications, the moduli space metric depends on geometric as well as topological data of the background through the warp factor and 2-form fluctuation potential $\eta_2$, both of which are determined through Poisson equations. 
Distinct to the methods of \cite{Martucci:2009sf,Martucci:2014ska,Martucci:2016pzt}, we arrive at these results without using SUSY, so our methods can apply to more general compactifications, as well.

The kinetic action for geometric K\"ahler moduli combined with their axionic partners (from $C_4$) draws on results from \cite{Frey:2008xw, Frey:2013bha} and appears in \eqref{actionKahler}; it agrees with the findings of \cite{Martucci:2016pzt}. 
We note that the 4D field space metric $C^{IJ}$ given via \eqref{eq:metricKahler4} depends on the background CY metric; for example, which representative of the cohomology is harmonic depends on the metric. Therefore, $C^{IJ}$ implicitly depends on the background values of geometric moduli which determine the background CY metric, so we promote it to the field space metric on the moduli space (since our analysis works at any point in moduli space). 

The moduli space metric for complex structure flat directions is reported here for the first time in the literature. Notably, we have not been able to rule out mixing between K\"ahler moduli and complex structure flat directions in the moduli space metric due to the shift in $G_3$, meaning that the K\"ahler and complex structure moduli spaces do not factorize. If that were to occur, that would be a surprising qualitative difference from the case of an unwarped product space CY compactification.

There are contemporary questions about what EFTs are allowed in UV-complete theories, so it is important to understand how to get the correct 4D EFT out of a given UV model. We have filled a gap by completing the derivation the 4D EFT for the K\"ahler moduli of GKP compactifications through the equations of motion and by presenting the moduli space metric of complex structure flat directions for the first time.

We conclude by briefly discussing several outlooks and potential applications of our results:
\begin{itemize}

\item
\textbf{\textit{Precision phenomenology:}} One future direction would be to develop computational tools capable of calculating the moduli space metric we have derived as the background CY metric is varied over the moduli space. This knowledge would allow for precision string phenomenology, even in the case of strong warping. Notably, there are recent advancements in numerically computing the metric on CYs and harmonic forms on them, e.g. see \cite{Mirjanic:2024gek, Hendi:2024yin, Halverson:2023ndu, Douglas:2020hpv, Ashmore:2020ujw, Cui:2019uhy, Ashmore:2019wzb, Anderson:2020hux, Douglas:2015aga}. These developments would play a crucial role in the suggested numerical computation of the moduli space metric.

\item
\textbf{\textit{Case of massive moduli:}} The dimensional reduction of the massive complex structure moduli also remains; a toy model and first steps appear in \cite{Frey:2006wv,Douglas:2007tu,Douglas:2008jx,Lust:2022xoq}. 
As seen here, deformations of the metric lead to shifts in the $G_3$ flux as well as the warp factor, both of which may affect the mass of stabilized moduli and have not yet been considered in detail. Determining the potential energy function of the 4D effective theory beyond perturbations around the background is an additional step.

\item
\textbf{\textit{Tadpole conjecture:}} The recently proposed tadpole conjecture \cite{Bena:2020xrh, Bena:2021wyr, Grana:2022dfw, Lust:2022mhk, Lust:2021xds, Plauschinn:2021hkp} suggests an upper bound on the number of stabilized complex structure moduli for compactifications on CY orientifolds with large Hodge number $h^{2,1}$. These tests had been conducted using the 4D superpotential (or period vector). For a given CY orientifold and flux quanta satisfying the D3 tadpole bound, determining which complex structure deformations satisfy the condition in \eqref{eq:ISDG3full} poses a counting problem (demonstrated in appendix \ref{app:CompStrFlat} in the simpler case of a toroidal orientifold which is not under the scope of the conjecture). 
We have seen that our approach is equivalent to using the 4D superpotential but offers an alternative formulation that may provide insight into proving the conjecture or finding counterexamples (if any) by scanning over flux vacua.
%This approach offers an alternative to using the 4D superpotential and could provide valuable insight into proving the conjecture or finding counterexamples (if any) by scanning over flux vacua.

\item
\textbf{\textit{Translation to 4D SUGRA:}}
In 4D SUGRA, the kinetic action and moduli space metric (as derived here) are controlled by the K\"ahler potential when written in terms of appropriately defined complex scalars (holomorphic variables). When restricted to the K\"ahler moduli sector, our results match the kinetic action found by \cite{Martucci:2016pzt} by alternate means. Therefore, the relation between 10D moduli and the 4D holomorphic variables and the K\"ahler potential in terms of those holomorphic variables are as given in \cite{Martucci:2016pzt}. The K\"ahler potential and 4D holomorphic variables for complex structure flat directions are not yet known. Nonetheless, as discussed at the end of section \ref{s:kineticcomplex}, consistency with the 10D ISD condition on the $G_3$ flux imposes stringent conditions on the K\"ahler potential of the complex structure moduli: either the first derivatives of the K\"ahler potential take the same functional form as for a product compactification, or the moduli space fails to factorize between K\"ahler and complex structure sectors but still maintains a no-scale structure. 
The first alternative occurs if the DeWolfe-Giddings K\"ahler potential \cite{DeWolfe:2002nn} can generate the flux compensator terms in the kinetic action for the complex structure flat directions.
Resolving this question is an important issue for future work.

\end{itemize}
%

% \paragraph{To do-}
% 1. Somewhere we need to mention that all the dot products use unwarped backgorund metric, except the ones in Appendix B which use the 10D metric. \rma{Now, added comment in Appendix A.}\\
% 2.

\acknowledgments
This work has been supported by the Natural Sciences and Engineering Research Council of Canada Discovery Grant program, grant 2020-00054.

\appendix

\section{Conventions}
\label{app:conventions}

In this appendix, we summarize the conventions used throughout the paper. We adopt mostly plus signature for the 10D and 4D metrics. Background fields are denoted either with a superscript or a subscript $(0)$. $x^\mu$ and $y^m$ respectively refer to the coordinates on the 4D spacetime and the 6D internal manifold, while $X$ is used to collectively stand for the coordinates $\{x,y\}$. Lowercase Greek indices $\mu, \nu, \dots$ run over the 4D spacetime coordinates, while lowercase Latin indices $m, n, \dots$ run over the internal $y$-coordinates. Uppercase Latin indices $M, N, \dots$ span the full 10D spacetime $X$.
The nomenclature $[i,j]$-form refers to a differential form on 10D spacetime with $i$ non-compact and $j$ compact indices.

$\tilde{d}$ denotes a differential with respect to $y$-coordinates, and covariant derivatives $\tilde{\nabla}_p$ are defined using the Christoffel connection for the unwarped background 6D metric $\tilde{g}^{(0)}_{pl}$. A tilde on upper indices indicates they are raised using the unwarped background 6D inverse metric $\tilde{g}^{(0)}{}^{pl}$. Similarly, $\hat{d}$ denotes a differential with respect to $x$-coordinates, and covariant derivatives $\hat{\nabla}_\mu$ are defined using the 4D Minkowski metric $\hat{\eta}_{\mu\nu}$, which coincide with the usual derivatives. Upper indices with hats are raised using the 4D Minkowski inverse metric $\hat{\eta}^{\mu\nu}$. In these notations, the 10D exterior derivative becomes $d=\hat{d}+\tilde{d}$. Additionally, for compactness, in a few instances indices are raised using the unwarped 6D inverse perturbed metric $\tilde{g}{}^{pl}$, and to indicate this a bar has been placed on such upper indices. Covariant derivatives with respect to the perturbed metric $\tilde{g}{}_{pl}$ are denoted by $\bar{\nabla}_p$.

The 4D and 6D volume forms respectively $\hat{\epsilon}$ and $\tilde{\epsilon}^{(0)}$ are defined by
\begin{align}
   \hat{\epsilon}\equiv\frac{1}{4!}\sqrt{-\hat{\eta}}\epsilon_{\mu_1\cdots\mu_4}dx^{\mu_1}\wedge \cdots\wedge dx^{\mu_4}\,, \quad \tilde{\epsilon}^{(0)}\equiv\frac{1}{6!}\sqrt{\tilde{g}^{(0)}}\epsilon_{m_1\cdots m_6}dy^{m_1}\wedge\cdots\wedge dy^{m_6}\,,
\end{align}
where $\epsilon_{\mu_1\cdots\mu_4},\epsilon_{m_1\cdots m_6}$ are respectively 4D and 6D Levi-Civita symbols. The Levi-Civita tensors (with metric determinant included) are therefore the components of these volume forms. For the Lorentzian spacetime, we choose $\epsilon_{0123}=+1$.

From time to time, we make use of the complex geometry on the Calabi-Yau manifold. We refer to $(i,j)$-form on the internal complex manifold as a differential form with $i$ holomorphic and $j$ antiholomorphic indices. The Dolbeault operators $\del \equiv dz^i \del_i$ and 
$\bar\del \equiv d\bar z^{\bar\imath}\del_{\bar\imath}$ act as exterior derivatives in the holomorphic and antiholomorphic directions respectively.

For notational simplicity, we omit the rank of a form when indices are explicitly shown. For example, $B_m$ and $B_{MN}$ represent the components of the compensator 1-form ($B_1$) and the Kalb-Ramond 2-form ($B_2$), respectively.

For any $p$-form $A_p$ on $n$-dimensional space(time) with metric $g_{kl}$ (and inverse $g^{kl}$), we take the definition of Hodge dual operation as
\begin{align}
  (\star A_p)_{k_1\dots k_{n-p}}=\frac{1}{p!}\sqrt{|g|}\epsilon_{k_1\dots k_{n-p}l_1\dots l_p}A^{l_1\dots l_p}\,,
\end{align}
where $\epsilon_{i_1\dots i_n}$ is the Levi-Civita symbol, and indices are raised with $g^{kl}$. For $p$-forms $A_p, B_p$, we define a symmetric dot product as
\begin{align}
  A_p\cdot B_p\equiv \textrm{sgn}(g)\ (-1)^{p(n-p)}\star(A_p\wedge\star B_p)=\frac{1}{p!}A_{k_1\dots k_p}B_{l_1\dots l_p}g^{k_1l_1}\cdots g^{k_pl_p}\,.
\end{align}
In complex coordinates, the dot product of a $(p,q)$ form with a $(q,p)$ form becomes
\begin{align}
A_{(p,q)}\cdot B_{(q,p)} = \frac{1}{p! q!} A_{i_1\cdots i_p\bar\jmath_1\cdots\bar\jmath_q} B_{a_1\cdots a_q\bar b_1\cdots \bar b_p} g^{i_1\bar b_1}\cdots g^{i_p\bar b_p}
g^{a_1\bar\jmath_1}\cdots g^{a_q\bar\jmath_q} .
\end{align}
Notably, all the dot products appearing in the paper use the unwarped background metric $\tilde{g}^{\0}_{mn}$, except for those in appendix \ref{app:IIBSUGRAeom}, which use the 10D metric.

The Hodge-de Rham Laplacian for the unwarped CY metric is 
\begin{equation}
\triangle \equiv -\t\star \t d\t\star \t d - \t d\t\star\t d\t\star\,,
\end{equation}
where the Hodge stars are associated with the metric $\t g_{mn}$.

Indices enclosed within square brackets or parentheses indicate antisymmetrization or symmetrization, respectively, defined as
\begin{align}
  A_{[k_1\cdots k_r]}\equiv\frac{1}{r!}\sum_{\sigma}\textrm{sgn}(\sigma)A_{k_{\sigma(1)}\cdots k_{\sigma(r)}}\,,\quad A_{(k_1\cdots k_r)}\equiv\frac{1}{r!}\sum_{\sigma}A_{k_{\sigma(1)}\cdots k_{\sigma(r)}}\,,
\end{align}
where $\sigma$ runs over all permutations of $(1, \dots, r)$, and $\text{sgn}(\sigma)$ represents the sign of the corresponding permutation.

\section{Type IIB supergravity equations of motion}
\label{app:IIBSUGRAeom}

The bosonic sector of the type IIB supergravity action in 10D Einstein frame is given by \cite{Polchinski:1998rr,Johnson:2000ch}
\be
  S_{\textrm{IIB}}=&\frac1{2\kappa^2} \bigg[ \int d^{10}X \sqrt{-g}\ R_g+\frac12\int d\Phi\wedge\star d\Phi + \frac12\int e^{-\Phi} H_3\wedge \star H_3\\
  &+\frac1{2}\int e^{2\Phi} F_1\wedge \star F_1 +\frac12\int e^{\Phi} \tilde{F}_3\wedge \star \tilde{F}_3+ \frac{1}4\int \tilde{F}_5\wedge \star \tilde{F}_5\bigg]\\
  &+\frac{1}{4\kappa^2}\int C_4\wedge H_3\wedge F_3\,,
\ee
where the NS-NS fields $g_{MN},B_{MN},\Phi$ are respectively the metric (with mostly plus signature), the Kalb-Ramond field and the dilaton, and the R-R form fields are $C_0$, $C_{MN}$, $C_{MNPQ}$. Their field strengths are given by
\be
  &H_3=dB_2\,,\quad F_1=dC_0\,,\quad F_3=dC_2\,,\quad F_5=dC_4\,,\quad \tilde{F}_3=F_3-C_0 H_3\,,\\
  &\tilde{F}_5=F_5-\frac12 C_2\wedge H_3+\frac12 B_2\wedge F_3\,.
\ee
Moreover, once the equations of motions are obtained, we shall impose a self-duality condition namely $\star\tilde{F}_5=\tilde{F}_5$ as a constraint from string theory. The Bianchi identities are given by
\be
  &dF_1=0\,,\quad dH_3=0\,,\quad d\tilde{F}_3=H_3\wedge F_1\,,\quad d\tilde{F}_5=H_3\wedge F_3\,. \label{eq:bianchi1}
\ee
Here, we list the equations of motion for the form fields and dilaton\footnote{Which have been simplified using the self-duality of $\tilde{F}_5$. And, the equation of motion for the $C_4$ field reduces to the Bianchi identity for $\tilde{F}_5$.}
\begin{align}
  &d(\star e^{-\Phi} H_3)-F^{(1)}\wedge\star e^{\Phi}\tilde{F}_3+\tilde{F}_5\wedge \tilde{F}_3=0\,,\quad d(\star e^{\Phi}\tilde{F}_3)+H_3\wedge \tilde{F}_5=0\,, \nn\\
  &d\star d\Phi+\frac12 e^{-\Phi} H_3\wedge \star H_3-e^{2\Phi}F_1\wedge \star F_1 -\frac12 e^{\Phi} \tilde{F}_3\wedge \star \tilde{F}_3=0\,, \nn\\
  &d(\star e^{2\Phi}F_1)+H_3\wedge \star e^{\Phi}\tilde{F}_3=0\,. \label{eq:eomforms1}
\end{align}
These can be rewritten in terms of axiodilaton $\tau$ and the 3-form flux $G_3$ as
\begin{align}
  &d\tilde{F}_5=\frac i2 \frac{G_3\wedge\bar{G}_3}{\textrm{Im}\tau}\,,\quad d\star G_3=-i \frac{d\tau\wedge\star \textrm{Re}G_3}{\textrm{Im}\tau}+i\tilde{F}_5\wedge G_3\,, \nn\\
  &d\star d\tau=-i\frac{d\tau\wedge\star d\tau}{\textrm{Im}\tau} -\frac{i}2 G_3\wedge\star G_3\,, \label{eq:eomforms2}\\
  &\tau\equiv C_0+i e^{-\Phi}\,,\quad G_3\equiv F_3-\tau H_3=\tilde{F}_3-i e^{-\Phi}H_3\,. \nn
\end{align}
And, the Bianchi identity for $G_3$ reads
\begin{align}
  dG_3=d\tau\wedge\frac{G_3-\bar{G}_3}{2i\textrm{Im}\tau}\,. \label{eq:bianchi2}
\end{align}
Now we list the Einstein field equations:
\begin{align}
  &G_{MN}=T^1_{MN}+T^3_{MN}+T^5_{MN}\,,\quad G_{MN}\equiv R_{MN}-\frac12 g_{MN} R_g\,, \nn\\
  &T^1_{MN}\equiv \frac1{2\left(\textrm{Im}\tau\right)^2}\partial_{(M}\tau\partial_{N)}\bar{\tau}-\frac1{4\left(\textrm{Im}\tau\right)^2} g_{MN}\ d\tau\cdot d\bar{\tau}\,,\nn\\
  &T^3_{MN}\equiv \frac1{2\cdot 2!\ \textrm{Im}\tau} G{}_{(M}{}^{PQ} \bar{G}{}_{N)PQ} - \frac1{4\textrm{Im}\tau} g_{MN}G_3\cdot\bar{G}_3\,, \nn\\
  &T^5_{MN}\equiv \frac{1}{4\cdot 4!} \tilde{F}_{MPQRS} \tilde{F}_N{}^{PQRS} -\frac{1}{8} g_{MN} \tilde{F}_5\cdot\tilde{F}_5=\frac{1}{4\cdot 4!} \tilde{F}_{MPQRS} \tilde{F}_N{}^{PQRS}\,, \label{eq:eomeinstein}
\end{align}
where in the last equality $\tilde{F}_5\cdot\tilde{F}_5$ vanishes due to the self-duality of $\tilde{F}_5$.

\section{Detailed fluctuations of terms in the equations of motion}
\label{app:FlucTermsEOMBian}

The fluctuation ansatzes for the metric and form fields are given in \eqref{eq:metricansatz} and \eqref{eq:formsansatz} in the main text. In this appendix, we report the fluctuations of various terms in the equations of motion and Bianchi identities \eqref{eq:eomforms2} - \eqref{eq:eomeinstein}, which result from those ansatzes.\footnote{In this appendix, for compactness, factors such as $e^{nA}, e^{m\Omega}, 1/\textrm{Im} \tau$ are retained in some expressions, and a few upper indices, which have a bar on them, are raised using the inverse metric $\tilde{g}{}^{pl}$. $\bar{\nabla}_p$ denotes a covariant derivative with respect to the perturbed metric $\tilde{g}{}_{pl}$. When expanded, such expressions generate order-zero and linear terms, and these linearized versions are used in section \ref{sec:Constr.dynEOM}. \label{fn:App.ConciseNotations}}

Let us start with the terms in Einstein filed equations. Up to linear order the components of the Einstein tensor become\footnote{Which have been simplified imposing conditions \eqref{eq:metricmodcons}.}
\begin{align}
  G_{\mu\nu}=& - 2e^{2\Omega+4A}\left(\bar{\nabla}^{\bar{2}}A-2\partial_m A\partial^{\bar{m}}A\right)\hat{\eta}_{\mu\nu} + 2\left(\partial^{\hat{2}}\Omega-2\partial^{\hat{2}}A-\frac{1}{2}e^{2\Omega+4A}\partial^{\hat{2}}u\tilde{\nabla}^{\tilde{m}}B_m\right)\hat{\eta}_{\mu\nu} \nn\\
  &\qquad - 2\left(\partial_{\mu}\partial_{\nu}\Omega-2\partial_{\mu}\partial_{\nu}A-\frac{1}{2}e^{2\Omega+4A}\partial_{\mu}\partial_{\nu}u\tilde{\nabla}^{\tilde{m}}B_m\right)\,,\nn\\
  G_{\mu m}=& - \frac{1}{2}e^{4A}\partial_{\mu}\partial_{m}e^{-4A} + \frac{1}{2} e^{4A}\partial^{\tilde{n}}e^{-4A}\partial_{\mu}u\delta\tilde{g}_{mn}-2e^{2\Omega+4A}\left(\tilde{\nabla}^{\tilde{2}}A-2\partial_{n}A\partial^{\tilde{n}}A\right)\partial_{\mu}uB_{m} \nn\\
  &\qquad + \frac{1}{2}e^{2\Omega+4A}\partial_{\mu}u\left(\tilde{\nabla}^{\tilde{n}}(\tilde{d}B_1)_{mn}+4\partial^{\tilde{n}}A(\tilde{d}B_1)_{mn})\right) +\frac{1}{2}\partial_\mu u\tilde{\nabla}^{\tilde{n}}\delta\tilde{g}_{mn}\,,\nn\\
  G_{mn}=& - \frac{1}{2}e^{-2\Omega-4A}\partial^{\hat{2}}u\delta\tilde{g}_{mn} - 8\partial_{m}A\partial_{n}A + 4\tilde{g}_{mn}\partial_{p}A\partial^{\bar{p}}A + e^{-2\Omega-4A}\left(3\partial^{\hat{2}}\Omega-2\partial^{\hat{2}}A\right)\tilde{g}_{mn} \nn\\
  &\qquad + \partial^{\hat{2}}u\tilde{\nabla}_{(m}B_{n)} + 4\partial^{\hat{2}}u\partial_{(
m}AB_{n)} - \partial^{\hat{2}}u(\tilde{\nabla}^{\tilde{p}}B_p)\tilde{g}_{mn} - 2\tilde{g}_{mn}\partial^{\tilde{p}}AB_{p}\partial^{\hat{2}}u\,.
\end{align}
And, the various contributions to the energy-momentum tensor defined in \eqref{eq:eomeinstein} up to linear order become
\begin{align}
  T^1_{MN}=&0\,, \nn\\
  T^3_{\mu\nu}=&\frac{1}{4}\hat{\eta}_{\mu\nu}\bigg[-\frac{e^{2\Omega+8A}}{\textrm{Im}\tau} G_3^{(0)}\cdot\bar{G}_3^{(0)} \nn\\
  &\quad\quad\quad+\frac{e^{2\Omega^{(0)}+8A^{(0)}}}{\textrm{Im}\tau^{(0)}} u(x) \left( \bar{G}_3^{(0)}\cdot W_3 - \bar{G}_3^{(0)}\cdot \chi_3 - \bar{\chi}_3\cdot G_3^{(0)} \right)\bigg]\,, \nn\\
  T^3_{\mu m}=&-\frac{1}{4}\partial_\mu u B_m \frac{e^{2\Omega^{(0)}+8A^{(0)}}}{\textrm{Im}\tau^{(0)}}G_3^{(0)}\cdot\bar{G}_3^{(0)} +\frac{ie^{4A^{(0)}}}{4\textrm{Im}\tau^{(0)}}\partial_{\mu}u \Big[\tilde{\star}^{(0)}(\eta_2\wedge\bar{G}_3^{(0)}-\bar{\eta}_2\wedge G_3^{(0)})\Big]_m\,, \nn\\
  T^3_{mn}=&0\,, \nn\\
  T^5_{\mu\nu}=&-4e^{2\Omega+4A}\left(\partial^{\bar{l}}A\partial_lA\right)\hat{\eta}_{\mu\nu}\,, \nn\\
  T^5_{\mu m}=&-4e^{2\Omega+4A}\left(\partial^{\tilde{l}}A\partial_lA\right)\partial_\mu uB_m +2e^{2\Omega+4A}\partial_\mu u (\tilde{d}B_1)_{mn}\partial^{\tilde{n}}A\,, \nn\\
  T^5_{mn}=&4\left(\partial^{\bar{l}}A\partial_lA\right)\tilde{g}_{mn}-8\partial_mA\partial_nA\,.
\end{align}
In above, $T^1_{MN}$ vanishes since the background value $\tau^{(0)}$ is a constant, and $T^3_{mn}$ vanishes owing to the ISD condition on the $[0,3]$ component of $G_3$ i.e. \eqref{eq:ISDG3}.

Now we linearize the terms that appear in the Bianchi identity of $\tilde{F}_5$ (which is same as $C_4$ equation of motion):\footnote{As indicated by the underbraces, the linear terms are grouped according to their $[i,j]$ components.}
\begin{align}
  d\tilde{F}_5 &= {\tilde{d}\tilde{\star}^{(0)}\tilde{d}e^{-4A^{(0)}(y)}} \nn \\
  &\qquad \underbrace{-4u(x)\tilde{d}\tilde{\star}^{(0)}\tilde{d}\left(e^{-4A^{(0)}}\delta A\right)-u(x)\tilde{d}\tilde{\star}^{(0)}V_1}_{[0,6]} \nn \\
  &\qquad \underbrace{-4\hat{d}u\wedge\tilde{\star}^{(0)}\tilde{d}\left(e^{-4A^{(0)}}\delta A\right)-\hat{d}u\wedge\tilde{\star}^{(0)}V_1+e^{2\Omega^{(0)}}\hat{d}u\wedge\tilde{d}\tilde{\star}^{(0)}\tilde{d}B_1}_{[1,5]} \nn \\
  &\qquad \underbrace{+e^{4\Omega^{(0)}}\hat{d}\hat{\star}\hat{d}u\wedge B_1\wedge \tilde{d}e^{4A^{(0)}(y)}-e^{4\Omega^{(0)}+4A^{(0)}}\hat{d}\hat{\star}\hat{d}u\wedge\tilde{d}B_1}_{[4,2]}\,.
\end{align}
\begin{align}
  &\frac{i}{2}\frac{G_3\wedge\bar{G}_3}{\textrm{Im}\tau}={\frac{i}{2}\frac{G_3^{(0)}\wedge\bar{G}_3^{(0)}}{\textrm{Im}\tau^{(0)}}} \nn \\
  &\qquad \underbrace{-\frac{i\textrm{Im}\delta\tau}{2(\textrm{Im}\tau^{(0)})^2}u(x)G_3^{(0)}\wedge\bar{G}_3^{(0)}+\frac{i}{2\textrm{Im}\tau^{(0)}}u(x)\left(G_3^{(0)}\wedge \bar{\chi}_3+\chi_3\wedge\bar{G}_3^{(0)}\right)}_{[0,6]} \nn \\
  &\qquad \underbrace{+\frac{i}{2\textrm{Im}\tau^{(0)}}\hat{d}u\wedge\left(\eta_2\wedge\bar{G}_3^{(0)}-G_3^{(0)}\wedge\bar{\eta}_2\right)}_{[1,5]}\,.
\end{align}
The first terms in above two equalities are of zeroth order, and they equate due to background equation of motion \eqref{eq:gkpbgwarpeqn}.

Now we linearize the terms that appear in the $G_3$ equation of motion:
\begin{align}
  d\star G_3&=e^{4\Omega^{(0)}}\hat{\epsilon}\wedge\tilde{d}(e^{4A^{(0)}}\tilde{\star}^{(0)}G_3^{(0)}) \nn\\
  &\qquad \underbrace{+4u e^{4\Omega^{(0)}}\hat{\epsilon}\wedge\left(\delta\Omega\tilde{d}(e^{4A^{(0)}}\tilde{\star}^{(0)}G_3^{(0)}) + \tilde{d}(e^{4A^{(0)}}\delta A\tilde{\star}^{(0)}G_3^{(0)})\right)}_{[4,4]} \nn\\
  &\qquad \underbrace{+ ue^{4\Omega^{(0)}}\hat{\epsilon}\wedge\tilde{d}e^{4A^{(0)}}\wedge\tilde{\star}^{(0)}(\chi_3-W_3) + ue^{4\Omega^{(0)}+4A^{(0)}}\hat{\epsilon}\wedge\tilde{d}\tilde{\star}^{(0)}(\chi_3-W_3)}_{[4,4]} \nn\\
  &\qquad \underbrace{+e^{4\Omega^{(0)}+4A^{(0)}}\hat{d}\hat{\star}\hat{d}u\wedge B_1\wedge\tilde{\star}^{(0)}G_3^{(0)} + e^{2\Omega^{(0)}}\hat{d}\hat{\star}\hat{d}u\wedge\tilde{\star}^{(0)}\eta_2}_{[4,4]} \nn\\
  &\qquad \underbrace{-e^{4\Omega^{(0)}}\hat{\star}\hat{d}u\wedge\tilde{d}(e^{4A^{(0)}}B_1\wedge\tilde{\star}^{(0)}G_3^{(0)}) - e^{2\Omega^{(0)}}\hat{\star}\hat{d}u\wedge\tilde{d}\tilde{\star}^{(0)}\eta_2}_{[3,5]}\,.
\end{align}
\begin{align}
  i\tilde{F}_5\wedge G_3&=ie^{4\Omega^{(0)}}\hat{\epsilon}\wedge\tilde{d}e^{4A^{(0)}}\wedge G_3^{(0)} \nn\\
  &\qquad \underbrace{+iue^{4\Omega^{(0)}}\hat{\epsilon}\wedge\tilde{d}e^{4A^{(0)}}\wedge\chi_3 + 4iue^{4\Omega^{(0)}}\hat{\epsilon}\wedge\left(\delta\Omega\tilde{d}e^{4A^{(0)}}+\tilde{d}(e^{4A^{(0)}}\delta A)\right)\wedge G_3^{(0)}}_{[4,4]} \nn\\
  &\qquad \underbrace{-ie^{4\Omega^{(0)}}\hat{\star}\hat{d}u\wedge\tilde{d}e^{4A^{(0)}}\wedge B_1\wedge G_3^{(0)} - ie^{4\Omega^{(0)}+4A^{(0)}}\hat{\star}\hat{d}u\wedge\tilde{d}B_1\wedge G_3^{(0)}}_{[3,5]}\,.
\end{align}
\begin{align}
  -i \frac{d\tau\wedge\star \textrm{Re}G_3}{\textrm{Im}\tau}=0\,.
\end{align}
The first terms in the first two equalities above are of zeroth order, and equating them gives $\tilde{d}G_3^{(0)}=0$ once we use the ISD property of the background flux $G_3^{(0)}$.

Now we linearize the terms that appear in the $\tau$ equation of motion:
\begin{align}
  &d\star d\tau=\delta\tau e^{2\Omega^{(0)}-4A^{(0)}}\hat{d}\hat{\star}\hat{d}u\wedge\tilde{\epsilon}^{(0)}\,, \\
  &-i\frac{d\tau\wedge\star d\tau}{\textrm{Im}\tau}=0\,,\\
  &-\frac{i}{2}G_3\wedge\star G_3=-\frac{i}{2}ue^{4\Omega^{(0)}+4A^{(0)}}\hat{\epsilon}\wedge G_3^{(0)}\wedge\left(\tilde{\star}^{(0)}(\chi_3-W_3)-i\chi_3\right)=0\,,
\end{align}
where in deriving the last equation we have used the ISD property of $G_3^{(0)}$ and condition \eqref{eq:ISDG3}.

As noted earlier, a few terms in preceding formulas have been presented in a concise form (see footnote \ref{fn:App.ConciseNotations}). Here, we present the linear expansion of one term that appears in dealing with Einstein field equations in the main text.
\begin{align}
  &\bar{\nabla}^{\bar{2}}A-4\partial_m A\partial^{\bar{m}}A=-\frac{e^{4A}}{4}\bar{\nabla}^{\bar{2}}e^{-4A}\,,\quad \bar{\nabla}^{\bar{2}}e^{-4A}=\tilde{\nabla}^{\tilde{2}}e^{-4A^{(0)}}+u\tilde{\nabla}^{\tilde{2}}\delta e^{-4A}-u\tilde{\nabla}^{\tilde{m}}V_{m}\,. \label{eq:PerturedLapA}
\end{align}
In deriving above, we have used the tracelessness of the metric fluctuations: $\tilde{g}^{mn}_{(0)}\delta \tilde{g}_{mn}=0$.

\section{Biscalar and bivector Green's functions}
\label{app:BiscaBivec}

In this appendix, we present the essential formulas related to biscalar and bivector Green's functions. For more details, refer to \cite{Cownden:2016hpf, Poisson:2011nh, Narlikar:1970x}.

We consider two points on the internal manifold, $y$ and $Y$, and use a slashed index to denote coordinates that transform under diffeomorphisms at the point $Y$, as opposed to the point $y$.

The biscalar and bivector Green's functions for the operator $\tilde{\nabla}^{\tilde{2}}$ w.r.t. the background metric $\tilde{g}^{(0)}_{mn}$ are respectively defined as
\begin{align}
  &\tilde{\nabla}^{\tilde{2}}\tilde{G}(y,Y)=\delta^6(y,Y)-\frac{1}{\tilde{V}_{(0)}}\,,\quad \tilde{\nabla}^{\tilde{2}}\tilde{G}_{\slashed{m}n}(y,Y)=\Lambda_{\slashed{m}n}\left(\delta^6(y,Y)-\frac{1}{\tilde{V}_{(0)}}\right)\,, \\
  &\int d^6Y \sqrt{\tilde{\slashed{g}}_{(0)}}\delta^6(y,Y)f(Y)=f(y)\quad \forall f\,. \nn
\end{align}
The parallel propagator $\Lambda_{\slashed{m}n}(y,Y)$ is defined so a covariantly constant vector field satisfies $A_{\slashed{m}}(Y) =\Lambda_{\slashed{m}n}(y,Y) A^{\t n}(y)$.
Since $\Lambda_{\slashed{m}n}$ is symmetric under the interchange of $\slashed{m}$ and $n$, $\tilde{G}_{\slashed{m}n}$ consequently inherits this symmetry. The functional dependencies are determined by the geodesic distance between $y$ and $Y$. The derivatives are related as
\begin{align}
  \tilde{\nabla}_{\slashed{m}}\tilde{G}(y,Y)=-\tilde{\nabla}_n \tilde{G}_{\slashed{m}}{}^{\tilde{n}}(y,Y)\,. \label{eq:BiscaBivecRel}
\end{align}

\section{Complex structure flat directions in \texorpdfstring{$T^6 / \mathbb{Z}_2$}{T6/Z2} compactifications}
\label{app:CompStrFlat}

In this appendix, we present explicit examples of flux backgrounds in $T^6 / \mathbb{Z}_2$ compactifications that illustrate the presence or absence of complex structure flat directions. These examples complement the discussion in section \ref{sec:GeoModComplex}. The flux choices are taken from \cite{Cicoli:2022vny}.\footnote{Based on a 4D superpotential analysis, \cite{Demirtas:2019sip} proposed a lemma providing conditions on 3-form flux quanta in order to have complex structure flat directions, later extended in \cite{Cicoli:2022vny}. Specifically, for $T^6/Z_2$ compactifications, these flux quanta were classified in \cite{Cicoli:2022vny} based on the number of complex structure flat directions they admit.}

We adopt the notation and basis for the harmonic forms on $T^6$ as presented in \cite{Kachru:2002he, Cicoli:2022vny}.\footnote{We do not present the basis here (for which we refer to the cited references) but directly provide the harmonic $G^{(0)}_3$ in the examples later.} We parameterize the real, periodic, right-handed coordinates on $T^6$ as $(y^1, \dots, y^6)$.\footnote{This differs from the conventions in \cite{Cicoli:2022vny}, where the coordinates are labeled as $(x^1, x^2, x^3, y^1, y^2, y^3)$. In our notation, $x$ refers to the non-compact 4D spacetime coordinates.} In the GKP backgrounds (also called flux vacua) under discussion, $T^6$ factorizes as $T^2 \times T^2 \times T^2$. The holomorphic 1-forms are given by $dz^i = dy^i + \tau^{(0)}_i dy^{i+3}$ for $i = 1, 2, 3$, where $\tau^{(0)}_i$ represent the background values of the complex structure moduli. This factorization results from a special class of flux choices considered in \cite{Cicoli:2022vny}, which can be relaxed to obtain potentially new flux vacua.

The integrals of $F_3,H_3$ over any 3-cycle are quantized (in units of $4\pi^2\alpha^\prime$). Thus, in the basis of harmonic 3-forms, $F_3,H_3$ fluxes can be specified by a set of integers maintaining tadpole bound. At the end, we present three such flux configurations, classified by the number of complex structure flat directions they allow. We also quote the corresponding value of $N_{\textrm{flux}}\equiv\int_{T^6}H_3\wedge F_3$ (in units of $16\pi^4\alpha^{\prime 2}$).\footnote{Note that, due to orientifolding, $N_{\textrm{flux}}/2$ enters the D3 tadpole condition.}

Given a flux configuration, we aim to identify the complex structure deformations of the metric for which equation \eqref{eq:ISDG3full} can be satisfied, thereby determining the flat directions. Before proceeding to specific examples, we first examine the complex structure deformations of the $T^6$ metric while preserving its determinant (hence its volume). For the GKP backgrounds under discussion, the metric on $T^6$ takes the off-diagonal form
\begin{align}
  \begin{pmatrix}
  0 & \tilde{g}^{(0)}_{i\bar{\jmath}}\\
  \tilde{g}^{(0)}_{\bar{\jmath}i} & 0
  \end{pmatrix}\,,
  \quad \tilde{g}^{(0)}_{i\bar{\jmath}}=\tilde{g}^{(0)}_{\bar{\jmath}i}=r_i\delta_{i\bar{\jmath}}\,.
\end{align}
Now, we vary the parameters as $ \tau^{(0)}_i \to \tau^{(0)}_i + u_i \delta \tau_i $, $ r_i \to r_i + \delta r_i $ in the line element $ ds_{T^6}^2 $, adjusting $ \delta r_i $ (in terms of $ u_i $ and their conjugates) to make the metric fluctuation traceless at linear order in $ u_i $ and $ \delta r_i $. This procedure eliminates linear terms proportional to $ dz^{i} d\bar{z}^{\bar{\jmath}} $ from the deformations of the line element, leaving only
\begin{align}
  ds_{T^6}^{\prime 2}=ds_{T^6}^2-i\sum_{i=1}^3\frac{r_i}{\textrm{Im}\tau_i^{(0)}}\left[\bar{u}_i\delta\bar\tau_i(dz^i)^2-u_i\delta\tau_i(d\bar{z}^i)^2\right]\,. \label{eq:T6CompStrMetDef}
\end{align}
Note that the parameters $\delta\tau_i$ correspond to basis $(2,1)$ forms $\xi^i$, so there is a single complex variable for each deformation, rather than separate $\delta^\pm\tau_i$ variables for $u_i$ and $\bar u_i$.
This allows us to extract the complex structure deformations $\delta^+\tilde{g}_{\bar{\imath}\bar{\jmath}},\delta^-\tilde{g}_{ij}$ (for each $u_i,\bar u_i$), which are then used to compute the components of $ W^{\pm}_3 $ via their definitions in \eqref{eq:pmDefs}. At this stage, since we do not know which combination(s) of the $(2,1)$ and $(1,2)$ harmonic basis forms correspond to complex structure flat directions, we retain all the parameters $ u_i, \bar{u}_i $ (collectively denoted by $\vec{u}$) in the metric fluctuations to compute $\delta_{\vec{u}} W_3 $. Additionally, we introduce a parameter $ u $ to vary the axiodilaton as $ \tau^{(0)} \to \tau^{(0)} + u \delta^+ \tau +\bar u\delta^-\tau$, which will later be determined in terms of $ u_i$ to ensure that we remain on the locus of ISD $ G_3 $ at linear order. This procedure will ultimately identify the flat direction(s) in our basis.

Some other generalities to note: In the flux vacua we consider below, $G^{(0)}_3$ is primitive $(2,1)$ harmonic form. This $G^{(0)}_3$, contracted with the metric deformations discussed above, can only generate the following nontrivial components: $W_{123}, W_{\bar{1}\bar{2}3}, W_{\bar{2}\bar{3}1}, W_{\bar{3}\bar{1}2}$. As illustrations, here we present\footnote{Here, we have: $\tilde{g}_{(0)}^{i\bar{\jmath}}=\delta^{i\bar{\jmath}}/r_i$.}
\begin{align}
  W_{123}&=\delta^-\tilde{g}_{11}G^{(0)}_{\bar{1}23}\tilde{g}_{(0)}^{1\bar{1}}+\delta^-\tilde{g}_{22}G^{(0)}_{\bar{2}31}\tilde{g}_{(0)}^{2\bar{2}}+\delta^-\tilde{g}_{33}G^{(0)}_{\bar{3}12}\tilde{g}_{(0)}^{3\bar{3}} \nn\\
  &=\frac{\bar{u}_1\delta\bar\tau_1}{i\textrm{Im}\tau^{(0)}_1}G^{(0)}_{\bar{1}23}+\frac{\bar{u}_2\delta\bar\tau_2}{i\textrm{Im}\tau^{(0)}_2}G^{(0)}_{\bar{2}31}+\frac{\bar{u}_3\delta\bar\tau_3}{i\textrm{Im}\tau^{(0)}_3}G^{(0)}_{\bar{3}12}\,, \nn\\
  W_{\bar{1}\bar{2}3}&=\delta^+\tilde{g}_{\bar{1}\bar{1}}G^{(0)}_{1\bar{2}3}\tilde{g}_{(0)}^{1\bar{1}}+\delta^+\tilde{g}_{\bar{2}\bar{2}}G^{(0)}_{23\bar{1}}\tilde{g}_{(0)}^{2\bar{2}} \nn\\
  &=\frac{iu_1\delta\tau_1}{\textrm{Im}\tau^{(0)}_1}G^{(0)}_{1\bar{2}3} +\frac{iu_2\delta\tau_2}{\textrm{Im}\tau^{(0)}_2}G^{(0)}_{23\bar{1}}\,.
\end{align}
With this setup in place, we now proceed to discuss the examples.

\paragraph{Case: one complex structure flat direction} The flux quanta (corresponding to $N_{\textrm{flux}}=24$) are such that the GKP background 3-form flux takes the value
\begin{align}
  G^{(0)}_3=\frac{1}{2i\textrm{Im}\tau^{(0)}}\left( 2 dz^1\wedge dz^2\wedge d\bar{z}^3-2 dz^1\wedge dz^3\wedge d\bar{z}^2-4 dz^2\wedge dz^3\wedge d\bar{z}^1 \right)\,,
\end{align}
which is a harmonic primitive $(2,1)$ form. In this example, the background values of the complex structure moduli $\tau^{(0)}_i$ coincide with the background value of the axiodilaton $\tau^{(0)}$. By requiring the $(3,0)$ component of $W_3$ to vanish and equating the $(1,2)$ components with the corresponding components of $i\delta_u\tau\bar{G}_3^{(0)}/\textrm{Im}\tau^{(0)}$, we obtain
\begin{align}
  &\bar{u}_2\delta\bar\tau_2 +\bar{u}_3 \delta\bar\tau_3=2\bar{u}_1 \delta\bar\tau_1\,, \nn\\
  &u_1\delta \tau_1+ \delta_u\tau=2 u_2\delta \tau_2\,, \nn\\
  & u_2\delta \tau_2+ u_3\delta \tau_3=2\delta_u\tau\,, \nn\\
  &u_1\delta \tau_1+\delta_u\tau=2 u_3\delta \tau_3\,.
\end{align}
Here, $u,\bar u$ parameterize the linear fluctuation $\delta_u\tau$, which are to be determined in terms of $u_i$ by solving the above equations. Because it only appears in equations with the $u_i$ variables and not the conjugate $\bar u_i$ variables, we can make a variable redefinition to set $\delta_u\tau=u\delta^+\tau$ with $\delta^-\tau=0$. 
The above system of equations admits only one non-trivial solution, given by $u_i\delta\tau_i=u\delta^+\tau$ for $i=1,2,3$. Since $u_i, u$ are variable parameters (and eventually become functions of spacetime), while $\delta\tau_i, \delta^+\tau$ are numerical coefficients, this implies that the $u_i$ are related to $u$, and $\delta\tau_i$ are related to $\delta^+\tau$, leaving the proportionality factors undetermined. 
Given that only the products $u_i\delta\tau_i$ and $u\delta^+\tau$ are relevant, we adopt a consistent choice of such factors so that the above solution reduces to 
\begin{align}
  u_i=u\,,\qquad \delta\tau_i=\delta^+\tau\,,\qquad i=1,2,3\,.
\end{align}
This result indicates that there is a single complex structure flat direction parameterized by $u$, which mixes with the axiodilaton. By substituting above conditions into the metric deformation \eqref{eq:T6CompStrMetDef}, we can read off the harmonic $(2,1)$ form $\xi$ and its conjugate (to be used in \eqref{eq:CompStrDefMet}) from the coefficients of $u,\bar{u}$. In our basis, this is given by the linear combination:\footnote{The proportionality constant can be determined by comparing \eqref{eq:CompStrDefMet} and \eqref{eq:T6CompStrMetDef}. The holomorphic 3-form on $T^6$ is taken to be $\tilde{\Omega}_3 = dz^1 \wedge dz^2 \wedge dz^3$, as used in \cite{Kachru:2002he, Cicoli:2022vny}.}
\begin{align}
  \xi\propto dz^1\wedge dz^2\wedge d\bar{z}^3+dz^2\wedge dz^3\wedge d\bar{z}^1+dz^3\wedge dz^1\wedge d\bar{z}^2\,.
\end{align}

\paragraph{Case: two complex structure flat directions}  The flux quanta (corresponding to $N_{\textrm{flux}}=8$) are such that the GKP background 3-form flux takes the value
\begin{align}
  G_3^{(0)}=\frac{1}{i\textrm{Im}\tau_3^{(0)}}\left(dz_1\wedge dz_2\wedge d\bar{z}_3-dz_2\wedge dz_3\wedge d\bar{z}_1\right)\,,
\end{align}
(a primitive harmonic $(2,1)$ form). In this example, the background values of the complex structure moduli $\tau^{(0)}_1$ and $\tau^{(0)}_3$ are equal, while $\tau^{(0)}_1$ coincides with the background value of the axiodilaton $\tau^{(0)}$. By requiring the $(3,0)$ component of $W_3$ to vanish and equating the $(1,2)$ components with the corresponding components of $i\delta_u\tau\bar{G}_3^{(0)}/\textrm{Im}\tau^{(0)}$, we obtain\footnote{The condition of vanishing of $(3,0)$ component coincides with the condition on $W_{\bar{3}\bar{1}2}$; similarly the conditions on $W_{\bar{1}\bar{2}3}$ and $W_{\bar{2}\bar{3}1}$ coincide.}
\begin{align}
  &u_2 \delta \tau_2=u \delta^+ \tau\,, \nn\\
  &u_3 \delta \tau_3=u_1 \delta \tau_1\,,
\end{align}
where $u$ parameterizes the linear fluctuation in $\tau$ (since we know the flat directions will be complex, we have already set $\delta^-\tau=0$ following the logic of the previous example). Since $u_i,u$ are variable parameters while $\delta\tau_i,\delta^+ \tau$ are numerical coefficients, this reduces to the conditions
\begin{align}
  u_1=u_3\,,\quad u_2=u\,,\quad \delta\tau_1=\delta\tau_3\,,\quad \delta\tau_2=\delta^+ \tau\,.
\end{align}
This result indicates that there are two complex structure flat directions parameterized by $u_1$ and $u_2$, where the latter mixes with the axiodilaton. By substituting above conditions into the metric deformation \eqref{eq:T6CompStrMetDef}, we can read off the harmonic $(2,1)$ forms $\xi^1,\xi^2$ and their conjugates (to be used in \eqref{eq:CompStrDefMet}) from the coefficients of $u_1,\bar{u}_1$ and $u_2,\bar{u}_2$ respectively. In our basis, these are given by the linear combinations:
\begin{align}
\xi^1\propto dz^3\wedge dz^1\wedge d\bar{z}^2\, ,\quad \xi^2\propto dz^1\wedge dz^2\wedge d\bar{z}^3+dz^2\wedge dz^3\wedge d\bar{z}^1\,.
\end{align}

\paragraph{Case: no complex structure flat directions} The flux quanta (corresponding to $N_{\textrm{flux}}=12$) are such that the GKP background 3-form flux takes the value
\begin{align}
  G_3^{(0)}=\frac{2e^{-\frac{i \pi }{6}}}{\sqrt{3}} \left( dz^1\wedge dz^2\wedge d\bar{z}^3- dz^1\wedge dz^3\wedge d\bar{z}^2+ dz^2\wedge dz^3\wedge d\bar{z}^1\right)\,,
\end{align}
which is a primitive harmonic $(2,1)$ form. In this example, the background values of the complex structure moduli $\tau^{(0)}_i$ coincide with the background value of the axiodilaton $\tau^{(0)}=e^{\frac{ 2 i\pi }{3}}$. By requiring the $(3,0)$ component of $W_3$ to vanish and equating the $(1,2)$ components with the corresponding components of $i\delta_u\tau\bar{G}_3^{(0)}/\textrm{Im}\tau^{(0)}$, we obtain the following equations:
\begin{align}
  &\bar{u}_1\delta \bar{\tau}_1 +\bar{u}_2\delta \bar{\tau}_2 +\bar{u}_3\delta \bar{\tau}_3 =0\,, \nn\\
  &u_1 \delta \tau_1 +u_2\delta \tau_2 =e^{\frac{i\pi}{3}}\delta_u\tau\,, \nn\\
  &u_2 \delta \tau_2 +u_3 \delta \tau_3 =e^{\frac{i\pi}{3}}\delta_u\tau\,, \nn\\
  &u_1\delta \tau_1 +u_3\delta \tau_3 =e^{\frac{i\pi}{3}}\delta_u\tau\,,
\end{align}
where $u$ parameterizes the linear fluctuation in $\tau$. This system of equations admits only the trivial solution, given by $u_i\delta\tau_i=\delta_u\tau=0$ for $i=1,2,3$. Since $u_i,u$ are variable parameters while $\delta\tau_i,\delta\tau$ are numerical coefficients, this reduces to the conditions
\begin{align}
  \delta\tau_i=\delta^+\tau=\delta^-\tau=0\,,\qquad i=1,2,3\,.
\end{align}
This result indicates that the given flux background does not admit any complex structure flat direction, i.e. all the complex structure moduli are stabilized.

Our results for the above flux backgrounds in $T^6/\mathbb{Z}_2$ compactification, which involve counting the complex structure flat directions and explicitly identifying them, are in perfect agreement with those obtained in \cite{Cicoli:2022vny} based on a 4D superpotential analysis.

\bibliography{biblio}
\bibliographystyle{JHEP}

\end{document}